\documentclass[twocolumn,twocolappendix]{aastex63}

\graphicspath{{./}{figures/}}
\usepackage[nolist]{acronym}
\newacro{adi}[ADI]{angular differential imaging}
\newacro{ao}[AO]{adaptive optics}
\newacro{aolp}[AOLP]{angle of linear polarization}
\newacro{bff}[BFF]{Best Factor Finding}
\newacro{charis}[CHARIS]{Coronagraphic High Angular Resolution Imaging Spectrograph}
\newacro{css}[CSS]{circumstellar signal}
\newacro{de}[DE]{differential evolution}
\newacro{disnmf}[DI-sNMF]{data imputation using sequential nonnegative matrix factorization}
\newacro{dpp}[DPP]{Data Processing Pipeline}
\newacro{drp}[DRP]{Data Reduction Pipeline}
\newacro{fov}[FOV]{field of view}
\newacro{fwhm}[FWHM]{full width at half maximum}
\newacro{gpi}[GPI]{Gemini Planet Imager}
\newacro{gto}[GTO]{Guaranteed Time Observations}
\newacro{hiciao}[HiCIAO]{High-Contrast Coronographic Imager for Adaptive Optics}
\newacro{hst}[HST]{Hubble Space Telescope}
\newacro{hwp}[HWP]{half-wave plate}
\newacro{i}[I]{total intensity}
\newacro{ir}[IR]{infrared}
\newacro{ifs}[IFS]{integral field spectrograph}
\newacro{irdis}[IRDIS]{InfraRed Dual-band Imager and Spectrograph}
\newacro{iwa}[IWA]{inner working angle}
\newacro{jwst}[JWST]{the James Webb Space Telescope}
\newacro{klip}[KLIP]{Karhunen-Lo\`{e}ve Image Projection}
\newacro{loci}[LOCI]{Locally Optimized Combination of Images}
\newacro{magaox}[MagAO-X]{MagAO-X}
\newacro{mloci}[MLOCI]{Matched LOCI}
\newacro{nir}[NIR]{near-infrared}
\newacro{nmf}[NMF]{Non-negative Matrix Factorization}
\newacro{opd}[OPD]{optical path difference}
\newacro{pca}[PCA]{principal component analysis}
\newacro{pdi}[PDI]{polarimetric differential imaging}
\newacro{pi}[PI]{polarized intensity}
\newacro{piaacmc}[PIAACMC]{phase-induced amplitude apodization complex mask coronagraph}
\newacro{ppd}[PPD]{protoplanetary disk}
\newacro{psf}[PSF]{point spread function}
\newacro{rdi}[RDI]{reference star differential imaging}
\newacro{sb}[SB]{surface brightness}
\newacro{scexao}[SCExAO]{Subaru  Coronagraphic  Extreme  Adaptive  Optics}
\newacro{sdi}[SDI]{spectral differential imaging}
\newacro{sed}[SED]{spectral energy distribution}
\newacro{snr}[SNR]{signal-to-noise ratio}
\newacro{snre}[SNRE]{signal-to-noise per resolution element}
\newacro{spf}[SPF]{scattering phase function}
\newacro{sphere}[SPHERE]{Spectro-Polarimetric High-contrast Exoplanet REsearch instrument}
\newacro{vampires}[VAMPIRES]{Visible Aperture Masking Polarimetric Imager for Resolved Exoplanetary Structures}
\newacro{vapp}[vAPP]{vector Apodizing Phase Plate}
\newacro{vlt}[VLT]{Very Large Telescope}
\usepackage{breqn}
\usepackage{hyperref}

\turnoffedit
\begin{document}
\title{JWST/NIRCam Coronagraphy of the Young Planet-hosting Debris Disk AU Microscopii}

\correspondingauthor{Kellen Lawson}
\email{kellenlawson@gmail.com}
\author[0000-0002-6964-8732]{Kellen Lawson}
\affiliation{NASA-Goddard Space Flight Center, Greenbelt, MD, USA}
\affiliation{NASA Postdoctoral Program Fellow}

\author[0000-0001-5347-7062]{Joshua E. Schlieder}
\affiliation{NASA-Goddard Space Flight Center, Greenbelt, MD, USA}

\author[0000-0002-0834-6140]{Jarron M. Leisenring}
\affiliation{Steward Observatory, University of Arizona, 933 N. Cherry Avenue, Tucson, AZ 85721, USA}

\author[0000-0002-7325-5990]{Ell Bogat}
\affiliation{NASA-Goddard Space Flight Center, Greenbelt, MD, USA}
\affiliation{Department of Astronomy, University of Maryland, College Park, MD 20782, USA}

\author[0000-0002-5627-5471]{Charles A. Beichman}
\affiliation{NASA Exoplanet Science Institute/IPAC, Jet Propulsion Laboratory, California Institute of Technology, 1200 E California Blvd, Pasadena, CA 91125, USA}

\author[0000-0001-5966-837X]{Geoffrey Bryden}
\affiliation{Jet Propulsion Laboratory, California Institute of Technology, Pasadena, CA, USA}

\author[0000-0001-8612-3236]{Andr\'as G\'asp\'ar}
\affiliation{Steward Observatory, University of Arizona, 933 N. Cherry Avenue, Tucson, AZ 85721, USA}

\author[0000-0001-5978-3247]{Tyler D. Groff}
\affiliation{NASA-Goddard Space Flight Center, Greenbelt, MD, USA}

\author[0000-0003-0241-8956]{Michael W. McElwain}
\affiliation{NASA-Goddard Space Flight Center, Greenbelt, MD, USA}

\author[0000-0003-1227-3084]{Michael R. Meyer}
\affiliation{Department of Astronomy, University of Michigan, 1085 S. University, Ann Arbor, MI 48109, USA}

\author[0000-0001-7139-2724]{Thomas Barclay}
\affiliation{NASA-Goddard Space Flight Center, Greenbelt, MD, USA}
\affiliation{University of Maryland, Baltimore County, Baltimore, MD 21250, USA}

\author[0000-0002-5335-0616]{Per Calissendorff}
\affiliation{Department of Astronomy, University of Michigan, 1085 S. University, Ann Arbor, MI 48109, USA}

\author[0000-0003-1863-4960]{Matthew De Furio}
\affiliation{Department of Astronomy, University of Michigan, 1085 S. University, Ann Arbor, MI 48109, USA}

\author[0000-0001-7591-2731]{Marie Ygouf}
\affiliation{Jet Propulsion Laboratory, California Institute of Technology, Pasadena, CA, USA}

\author[0000-0001-9353-2724]{Anthony Boccaletti}
\affiliation{LESIA, Observatoire de Paris, Université PSL, CNRS, Sorbonne Université, Université de Paris Cité, 5 place Jules Janssen, 92195 Meudon, France}

\author[0000-0002-8963-8056]{Thomas P. Greene}
\affiliation{NASA Ames Research Center, MS 245-6, Moffett Field, CA 94035, USA}

\author{John Krist}
\affiliation{Jet Propulsion Laboratory, 4800 Oak Grove Drive M/S 183-900, Pasadena, CA 91109, USA}

\author[0000-0002-8864-1667]{Peter Plavchan}
\affiliation{George Mason University Department of Physics \& Astronomy, 4400 University Drive, MS 3F3, Fairfax, VA 22030, USA}

\author[0000-0002-7893-6170]{Marcia J. Rieke}
\affiliation{Steward Observatory, University of Arizona, 933 N. Cherry Avenue, Tucson, AZ 85721, USA}

\author[0000-0002-6730-5410]{Thomas L. Roellig}
\affiliation{NASA Ames Research Center, MS 245-6, Moffett Field, CA 94035, USA}

\author[0000-0003-2434-5225]{John Stansberry}
\affiliation{Space Telescope Science Institute, 3700 San Martin Drive, Baltimore, MD 21218, USA}

\author[0000-0001-9209-1808]{John P. Wisniewski}
\affiliation{NASA Headquarters, 300 Hidden Figures Way SW, Washington, DC 20546, USA}

\author[0000-0002-6395-4296]{Erick T. Young}\affiliation{Universities Space Research Association, 425 3rd Street SW, Suite 950, Washington DC 20024, USA}

\submitjournal{\aj}
\accepted{2023 August 01}

\begin{abstract}
High-contrast imaging of debris disk systems permits us to assess the composition \edit1{and size distribution} of circumstellar dust, to probe recent dynamical histories, and to directly detect and characterize embedded exoplanets. Observations of these systems in the infrared beyond 2--3 $\, \micron$ promise access to both extremely favorable planet contrasts and numerous scattered-light spectral features — but have typically been inhibited by the brightness of the sky at these wavelengths. We present coronagraphy of the AU Microscopii (AU Mic) system using JWST's Near Infrared Camera (NIRCam) in two filters spanning 3--5$\, \micron$. These data provide the first images of the system's famous debris disk at these wavelengths and permit additional constraints on its properties and morphology. Conducting a deep search for companions in these data, we do not identify any compelling candidates. However, with sensitivity sufficient to recover planets as small as $\sim 0.1$ Jupiter masses beyond $\sim 2\arcsec$ ($\sim 20$ au) with $5\sigma$ confidence, these data place significant constraints on any massive companions that might still remain at large separations and provide additional context for the compact, multi-planet system orbiting very close-in. The observations presented here highlight NIRCam's unique capabilities for probing similar disks in this largely unexplored wavelength range, and provide the deepest direct imaging constraints on wide-orbit giant planets in this very well studied benchmark system.
\end{abstract}

\keywords{}

\section{Introduction}\acresetall

\edit1{Compact multi-planet systems are ubiquitous around the galaxy's lowest mass stars \citep{Dressing2015, Hardegree2019}, so their formation and evolution, and the impact of giant, outer companions, is of great interest. This is particularly true in the context of galactic planet demographics \citep{Gaudi2021, Barclay2017, Meyer2018}, exoplanet habitability \citep{Clement2022}, and as analogs to solar system formation \citep{Childs2019}. The presence or absence of giant planets in the outskirts of M dwarf systems can have a substantial impact on the delivery of the volatiles necessary for habitability to terrestrial planets deep in their interiors \citep{Alibert2017, Bitsch2019, Schlecker2021, Clement2022}. To date, such planets} remain generally elusive and particularly difficult to discover \citep{Bowler2015, Lannier2016}. Though microlensing surveys provide evidence for a significant population of sub-Jupiters at $\sim$1-10 au separations \citep{Shvartzvald2016, Suzuki2016}, direct imaging studies have previously been limited in their ability to detect planets in this mass regime. Early work by \citet{Beichman2010} showed that the wavelength coverage and predicted contrast of JWST's high contrast imaging modes would be ideal to detect such planets around nearby M dwarfs. These capabilities would open an unprecedented region of the mass-separation parameter space for exoplanets and provide unique insights into the population of planets in the outskirts of M dwarf systems. These predictions led to a deeper investigation into the optimal targets for coronagraphy from JWST's Near-infrared Camera \citep[NIRCam;][]{Rieke2005,Rieke2023} and new analyses of JWST's high contrast imaging capabilities in the context of M dwarfs using updated performance predictions and planet models \citep{Schlieder2015, Brande2020, Carter2022}. These studies revealed that particularly nearby and young M dwarfs provide access to the lowest mass planets at the closest separations — with the potential to push below a Saturn mass at $\lesssim$10 au. 

AU Microscopii (AU Mic, HD 197481, GJ 803) is a very nearby \citep[$d=9.714\pm0.002$ pc;][]{GaiaDR32022}, bright ($V$=8.6, $K_s$=4.5) M dwarf. Spectrophotometric observations from the late 1950's provided the first estimates of an early-M type \citep{Vyssotsky1956}. \cite{Keenan1989} measured a spectral type of M1V using modern spectroscopic techniques. Indications of strong magnetic activity and a pre-main sequence age were emerging as early as the 1970s, with the detection of chromospheric emission lines, flares, and spot driven variability  \citep{Wilson1970, Kunkel1973, Torres1973}. 

\cite{Barrado1999} ultimately confirmed the young age of AU Mic by a combination of kinematic association with the young A-type star $\beta$ Pictoris ($\beta$ Pic), elevated HR diagram position, and strong activity. These diagnostics indicated an age of 20$\pm$10 Myr. Further follow-up revealed that AU Mic and many other nearby, young M dwarfs, are members of the $\beta$ Pic moving group ($\beta$PMG) kinematic association \citep[e.g.,][]{Zuckerman2001, Schlieder2010, Schlieder2012, Shkolnik2017, Gagne2018}. The age of the $\beta$PMG has been determined to be $\approx$20-25 Myr using numerous techniques that include color-magnitude diagram and isochrone analyses \citep[e.g.,][]{Bell2015}, lithium depletion boundary measurements \citep[e.g.,][]{Binks2014}, binary dynamical mass constraints \citep[e.g.,][]{Nielsen2016}, and kinematic traceback estimates 
\citep[e.g.,][see Table \edit1{6} of \citealt{MiretRoig2020} and references therein for a literature summary of $\beta$PMG age estimates]{MiretRoig2020, Couture2023}. \citet{Bell2015} provide an isochronal age estimate of 24$\pm$3 Myr for the $\beta$PMG, which we adopt as the age of AU Mic hereafter. 

The Infrared Astronomical Satellite (IRAS) provided the first evidence of a debris disk around AU Mic via a $60 \, \micron$ excess \citep{Mathioudakis1991,Mathioudakis1993,Mullan1989}.
Subsequent reanalysis of IRAS data by \citet{Song2002} confirmed the \ac{ir} excess at $60 \, \micron$ and refined the model of the \ac{sed}. At $850 \, \micron$, the system was detected as a point source — indicating the presence of an unresolved population of cold ($\sim$40 K) dust \citep{Liu2004a}. 

 Around this time, visible band coronagraphy provided the first resolved images of the nearly edge-on disk — extending from the inner edge of the coronagraph at $\sim$50 au to separations of $\sim$210 au in reflected light \citep{Kalas2004}. Follow-up observations using \ac{nir} coronagraphy revealed the disk extending to smaller separations and identified sub-structure and asymmetries between the southeast and northwest sides \citep{Liu2004b}. Over the nearly two decades that followed, continuous high-contrast and high-resolution imaging spanning optical to millimeter wavelengths \citep[e.g.][]{Fitzgerald2007, MacGregor2013} have made the AU Mic disk one of the best studied. Notable features include a planetesimal belt at $\sim$40 au \citep{Augereau2006}, an extended halo \citep{Matthews2015}, and dynamical changes in the disk, including a color change \citep{Lomax2018} and fast-moving sub-structures \citep{Boccaletti2015, Sezestre2017, Boccaletti2018, Wisniewski2019, Grady2020}. 

More recently, a compact system of transiting exoplanets — well within the inner edge of its planetesimal belt — has been identified \citep{Plavchan2020, Martioli2021, Gilbert2022, Wittrock2023}. Two of the planets, AU Mic b and c, have radii and masses comparable to Neptune and Uranus \citep{Klein2021, Cale2021, Zicher2022}, but have orbital periods of only 8.46 and 18.86 days (semimajor axes of 0.0645 and 0.1101 au respectively). These periods lead to a resonance that drives transit timing variations (TTVs) on the order of minutes \citep{Wittrock2022}, providing a deeper characterization of the system and evidence for additional perturbers. \citet{Wittrock2023} identify a third planet, AU Mic d, via TTVs, with a model-favored period of 12.74 days and a mass comparable to that of the Earth. The periods of the AU Mic planets are close to a 4:6:9 resonance and provide clues to planet formation and evolution in the deep interior of this young system \citep{Wittrock2023}. 

A selection of M dwarf targets is currently being studied through a JWST NIRCam Guaranteed Time Observing (GTO) program. This program, GTO 1184 (PI J. Schlieder)\footnote{\href{https://www.stsci.edu/jwst/science-execution/program-information.html?id=1184}{GTO 1184 - A NIRCam Coronagraphic Imaging Survey of Nearby Young M Dwarfs}} targets 9 of the closest, youngest M dwarfs using NIRCam coronagraphy in two filters spanning 3-5 $\mu$m. AU Mic's four-fold combination of proximity, young age, dynamically active debris disk, and exoplanet system place it among the richest laboratories to study the formation and evolution of planetary systems around low-mass stars. Moreover, analysis in \citet{Sezestre2017} explored the origin of AU Mic's fast-moving disk features and found plausible scenarios involving an orbiting companion with semimajor axis between roughly 5 and 25 au. For these reasons, AU Mic was considered a prime target for GTO 1184. 

In this study, we present the results of JWST/NIRCam observations of AU Mic from GTO 1184. We report conclusive first detections of the famous debris disk at both $3.6 \, \micron$ and $4.4 \, \micron$, and spanning separations from $\sim 0\farcs3$ to $\sim 5\arcsec$ (2.9 to 49 au). With these data, we conduct a deep search for companions — effectively probing companion masses well below that of Saturn — and provide preliminary analyses of the disk itself. A follow-up study will perform further analysis focusing on the disk.

\section{Observations}\label{sec:observations}

Rather than conducting dedicated reference observations to accompany each science target, GTO 1184 improves survey efficiency by adopting a self-referencing strategy for \ac{rdi}. In this strategy, \ac{rdi} is performed for each target using the suite of other targets' images as a reference library (with consideration for any off-axis sources that may be present in the library). All observations used the \texttt{MASK335R} coronagraph mask \citep{Krist2009}, which has an \ac{iwa}\footnote{\edit1{Where the \ac{iwa} is defined as the angular separation at which coronagraph transmission reaches 50\%}} of $0\farcs63$. 

Observations were conducted using two filters from the NIRCam long wavelength (LW) channel \edit1{(average pixel scale of 63 mas$/$pixel)}: F356W ($\lambda_{pivot} = 3.563 \, \micron$, $\Delta \lambda = 0.787 \, \micron$) and F444W ($\lambda_{pivot} = 4.421 \, \micron$, $\Delta \lambda = 1.024 \, \micron$)\citep{Rieke2005}. \edit1{Using the \texttt{SUB320} subarray mode, each integration in the LW channel produces an image of 320$\times$320 pixels (20$\times$20$\arcsec$)}.
\edit1{The F356W and F444W} filters were selected primarily to maximize sensitivity to planets and to enable rejection of background sources via color analyses. \edit1{For NIRCam's non-coronagraphic imaging mode, wavelengths shorter than $\sim$$4 \, \micron$ in the LW channel produce \acp{psf} with \ac{fwhm} smaller than two pixels and are thus considered undersampled. However, for coronagraphy, the Lyot stop that is paired with the round coronagraph masks reduces the effective telescope diameter from 6.5 meters to 5.2 meters. This results in \ac{psf} \acp{fwhm} of approximately 2.29 pixels ($0\farcs14$) and 2.84 pixels ($0\farcs18$) for F356W and F444W, respectively. As such, neither filter's \ac{psf} is undersampled when used for coronagraphy.}

All targets were observed at two roll angles separated by $\sim 10\degr$, with the exception of TYC 5899 (for which only a single roll was executed on the indicated date due to a target acquisition failure). To accommodate its relative brightness, observations of AU Mic used the \texttt{SHALLOW2} readout pattern, while all other observations used \texttt{MEDIUM8}. The signal-to-noise ratio in the AU Mic integrations is somewhat higher than for the reference integrations (by a factor of $\sim 2$ compared to HIP 17695 and TYC 5899 — the closest targets in brightness). However, when the reference integrations are combined during \ac{rdi} (see Section \ref{sec:psfsub}), the SNR for the resulting models of the stellar diffraction pattern in each science integration is much more comparable. Nevertheless, this should be expected to slightly diminish the sensitivity in the background-limited regime ($r \gtrsim 2\arcsec$) compared to a conventional \ac{rdi} sequence in which the reference observations are tuned to achieve comparable SNR.

A summary of the GTO 1184 JWST/NIRCam observations used in this study, including observation dates and instrument settings, is provided in Table \ref{tab:observations}. Targets utilized as references were found to be free of extended circumstellar emission, with any nearby sources in the reference images being accounted for as described in Section \ref{sec:psfsub}. The spectral type mismatch between AU Mic and the reference targets is not expected to significantly impact sensitivity \citep[e.g.,][]{Girard2022}.
Other available GTO 1184 observations excluded from Table \ref{tab:observations} were not ultimately used as they were not found to improve \ac{rdi} subtraction for AU Mic. A full description of the GTO 1184 program and the survey results will be presented in a forthcoming publication.

\begin{deluxetable*}{@{\extracolsep{5pt}}ccccccccc}
    \tablewidth{0pt}
    \tablecaption{JWST/NIRCam Observations}
    \tablehead{
    \colhead{} & \colhead{} & \colhead{} & \colhead{} & \colhead{} & \multicolumn{2}{c}{F356W} & \multicolumn{2}{c}{F444W} \\
    \cline{6-7}
    \cline{8-9}
    \colhead{Prop. ID\tablenotemark{a}} & \colhead{Spec. Type} & \colhead{W1$\,$(mag)\tablenotemark{b}} & \colhead{W2$\,$(mag)\tablenotemark{b}} & \colhead{Obs. Date} & \colhead{$\rm N_{int}$\tablenotemark{c}} & \colhead{Exp. Time\tablenotemark{d}} & \colhead{$\rm N_{int}$\tablenotemark{c}} & \colhead{Exp. Time\tablenotemark{d}}
    }
    \startdata
    V-AU-MIC & M1 & 4.45 & 4.01 & 2022 Oct 3 & 34 & 1708 & 70 & 3517 \\
    HIP 17695 & M3 & 6.81 & 6.66 & 2022 Oct 3 & 16 & 1676 & 34 & 3562 \\
    G-7-34 & M4 & 8.01 & 7.81 & 2022 Oct 3 & 16 & 1676 & 34 & 3562 \\
    TYC 5899\tablenotemark{e} & M3 & 6.78 & 6.58 & 2022 Oct 3 & 8 & 838 & 17 & 1781 \\
    2MJ0443\tablenotemark{f} & M9 & 10.83 & 10.48 & 2022 Sep 6 & 16 & 1676 & 34 & 3562 \\
    LP-944-20 & M9 &  9.13 & 8.80 & 2022 Sep 6 & 16 & 1676 & 34 & 3562 \\
    \enddata
    \tablecomments{A summary of the science and reference targets and their observations used in this study. All targets except AU Mic (``V-AU-MIC") were used as reference targets. \tablenotetext{a}{The identifier used for each target in the GTO 1184 proposal.} \tablenotetext{b}{ALLWISE photometry; \citet{Wright2010}} \tablenotetext{c}{The total number of integrations for the corresponding filter; every integration had 10 groups.} \tablenotetext{d}{The total effective exposure time over all integrations in units of seconds.} \tablenotetext{e}{TYC 5899-193-1} \tablenotetext{f}{2MASS J0443376+000205}}
    \label{tab:observations}
\end{deluxetable*}

\section{Data Reduction and Post-Processing}
To reduce the data, we make use of the \texttt{spaceKLIP} package \citep[][which, in turn, uses the \texttt{JWST Pipeline}; \citealt{Bushouse2022}]{Kammerer2022} and largely follow the procedure of \citet{Carter2022}, summarized hereafter, with some exceptions as indicated.

Beginning from Stage 0 products (\textit{*uncal.fits}), we process the data to Stage 1 (\textit{*rateints.fits}) using the \texttt{rampfit} step of \texttt{spaceKLIP}. Following \citet{Carter2022}, we a) use the updated reference pixel definition described therein, b) skip the dark current subtraction step to avoid the negative effects of the low-quality calibration data that are currently available, and c) adopt a jump detection threshold of 5.

We then process the resulting Stage 1 products to Stage 2 (\textit{*calints.fits}) using the \texttt{imgprocess} step from \texttt{spaceKLIP}. However, we found that the included pixel cleaning procedures from both \texttt{spaceKLIP} and the \texttt{JWST Pipeline} left numerous problematic pixels uncorrected. The apparent differences in performance between application to the data of \citet{Carter2022} and our data may result from differences in the observing strategies — e.g., with \citet{Carter2022} observing a dedicated reference target with a small grid dither strategy versus our self-referenced survey lacking dedicated reference targets and dithers. Instead, we executed the \texttt{imgprocess} step without pixel cleaning, and performed subsequent pixel cleaning using the procedures outlined hereafter.

\subsection{Outlier Rejection}
We rejected outlier pixels based on three criteria, initially replacing offending pixels with NaN values in each case:

\begin{enumerate}
    \item pixels falling arbitrarily close to zero (absolute values less than 1e-8)
    \item pixels having data quality (DQ) flag values above 100 (e.g., flagged cosmic ray events)
    \item pixels falling more than 7.5 median absolute deviations (MADs) from the median among the values of a given pixel across all integrations of a single roll angle (e.g., unflagged cosmic ray events)
\end{enumerate}

For each criterion, the numerical threshold was set ad hoc based on visual inspection of the data. After this procedure, we manually masked an additional 28 seemingly discrepant pixels that were not identified by the criteria above. Generally, these were pixels that were persistently much brighter or fainter than neighboring pixels (i.e., spatial outliers rather than temporal outliers). We remark that it may be feasible to automatically reject these pixels with careful application of something akin to a tophat filter. However, in limited testing, we were unable to identify a process that would not occasionally mask bright diffraction speckles as well. Once this process was finished, we replaced each masked pixel with the median of values within a 5$\times$5 pixel window.

\subsection{Image Registration}\label{sec:imreg}
To identify the position of the occulted star in the data (and thus the misalignment between the star and the coronagraph), we largely follow the procedure of \citet{Carter2022}. Since any pointing changes between integrations are expected to be negligible \citep[$\sim 1 \,$mas RMS][]{Rigby2023}, we median-combine the integrations of each exposure for determining offsets before then applying corresponding shifts to the individual integrations.

To begin, we generate a synthetic coronagraphic image of a perfectly centered star using \texttt{WebbPSF} with the closest \ac{opd} map to the observations (\texttt{O2022100401-NRCA3\_FP1-1.fits}; measured approximately one day after the collection of the AU Mic data). We then use cross correlation with this synthetic image to determine the position of the star in the first exposure of the observations of HIP 17695 (the target observed nearest in time to AU Mic) — hereafter referred to as the centering reference —  using the \texttt{JWST Pipeline} image registration procedure (\texttt{imageregistration.align\_array}). We chose to use HIP 17695 for this purpose to avoid any impact from AU Mic's bright disk, which is faintly visible in the raw data. We then cross correlated each additional exposure with the centering reference to determine the relative offsets between exposures.

Throughout this procedure, we consider only the pixels within a coronagraph-centered annulus spanning $12 < r < 32$ pixels. While the methodology of \citet{Carter2022} uses an 11$\times$11 pixel coronagraph-centered box for relative centering, our data span a much larger temporal baseline — making small-separation changes to the diffraction pattern more likely. Within the aforementioned annular region considered for relative centering, centering for the AU Mic images explicitly excludes a rectangular region approximately aligned with the disk's major axis (adjusted for the parallactic angle of the exposure) — assuming a disk position angle of $128\fdg48$ \citep{Vizgan2022} and a width of 12 pixels. Once these offsets are computed, we shift all of the integrations such that the stars are aligned with the reference pixel for the NIRCam long wavelength (LW) target acquisition (TA) filter, F335M. This accounts for a recently identified filter-dependent offset for NIRCam, which shifts the entire scene relative to the TA filter (\textit{J. Leisenring, private communication}). In detector pixels, these offsets are $[x,y] = [0.751, -0.121]$ for F356W, and $[x,y] = [0.157, -0.224]$ for F444W. 

This procedure is carried out separately for each filter. For the two rolls of the F356W observations of AU Mic, we find $[x,y]$ offsets from the coronagraph center to the star of [$30$ mas, $-6$ mas] and [$18$ mas, $-6$ mas]. For F444W, we find offsets of [$24$ mas, $-6$ mas] and [$12$ mas, $-5$ mas].

\subsection{Reference Star Differential Imaging}\label{sec:psfsub}
In a pure ``classical" \ac{rdi} procedure, the pattern of diffracted starlight in the science target's sequence would typically be modeled by taking the median of the sequence of reference images and then scaling the result to match the brightness of the target based on prior knowledge of the two stars' fluxes. With consideration for the variability of the targets in our data and the lack of precise knowledge regarding the stellar fluxes in these bandpasses, such a procedure is problematic. Instead, we estimate the ratio of the stellar fluxes empirically by median combining the science target and reference sequences and then determining the scaling factor that minimizes the squared residuals between the two median images within a region that explicitly excludes the vicinity of AU Mic's disk (using the same region defined in Section \ref{sec:imreg}). We then multiply the median of the reference images by this scaling factor and subtract it from the target sequence. For this purpose, the reference sequence includes only the images of HIP 17695 --- the star in our sample whose observations were closest in time and whose spectral type is the closest match for that of AU Mic. In addition to this classical RDI reduction, we also carry out an RDI / \ac{klip} reduction as implemented in \texttt{SpaceKLIP} \citep{Kammerer2022} --- utilizing the full frame and again using the images of HIP 17695 as the reference data\footnote{The RDI/KLIP reduction with \texttt{SpaceKLIP} performs the nominal centering strategy described in \citet{Carter2022} as part of the starlight subtraction procedure, but otherwise uses the same outlier-corrected (``cleaned") data as the other reductions.}. Since the NIRCam diffraction pattern changes based on the alignment between the target star and the coronagraph, techniques like KLIP are expected to more effectively eliminate starlight by combining multiple reference images such that residual flux in the science image is minimized.

To better mitigate stellar residuals at small separations, we additionally apply a \acl{loci} \citep[\acs{loci}; ][]{Lafreniere2007} \ac{rdi} procedure utilizing just two optimization/subtraction regions and two separate sets of reference images. The inner optimization region covers $r < 20$ pixels ($1\farcs26$), with subtraction being performed over the same region, and incorporates reference  images from the sequences for HIP 17695, 2MJ0443, G-7-34, LP-944-20, and TYC 5899. Meanwhile, the outer optimization region spans $10 < r < 35$ ($0\farcs63 - 2\farcs21$; from roughly the \ac{iwa} to where a background source begins to infringe for AU Mic), with the corresponding subtraction region covering $r \ge 20$ pixels ($1\farcs26$), and includes only reference images from HIP 17695 and G-7-34, which were found to be free of off-axis sources within the optimization region. This strategy allows us to benefit from a more diverse selection of reference images at small separations --- where differences in the wavefront or coronagraph alignment change the coronagraphic image more significantly --- without being affected by the bright off-axis sources present at larger separations for many of the survey targets.

With KLIP or LOCI, a model of the starlight in an image is constructed from some combination of reference images to minimize the residuals with a science image. When \ac{css} is present in the science image, these techniques effectively identify the combination of reference images that best nulls the entire scene, including both stellar and circumstellar signal. Unlike for classical RDI, this results in a model of the starlight that is systematically brighter than it should be and that ultimately suppresses the throughput of all circumstellar flux in the science data — an effect referred to as ``oversubtraction''. The significance of oversubtraction depends on the relation between the spatial distributions of the circumstellar and stellar flux such that it is often negligible for point sources. However, the effect is often severe for extended sources — suppressing a significant fraction of flux while also introducing color offsets and inducing or obfuscating morphological features.

To attain image products free of significant oversubtraction, we also carry out model-constrained RDI (MCRDI), as described in \citet{Lawson2022}, in which a model of the \ac{css} is optimized alongside the model of the starlight. By subtracting the \ac{css} estimate from the data during optimization of the stellar model, the contribution to the stellar model resulting from the projection of the underlying disk signal onto the reference images (or reference eigenimages in the case of KLIP), the cause of \ac{rdi} oversubtraction, can be eliminated. Effectively, this techniques seeks the models of the stellar and circumstellar light that best explain the data — then subtract the resulting stellar model to isolate the circumstellar flux. A description of the utilized synthetic disk model, its parameters, and the optimization procedure are provided in Appendix \ref{app:mcrdi_models}. With the exception of the model-based ``constraint", this reduction is identical to the LOCI RDI reduction described previously. 

\begin{figure*}
\centering
\includegraphics[width=0.9\textwidth]{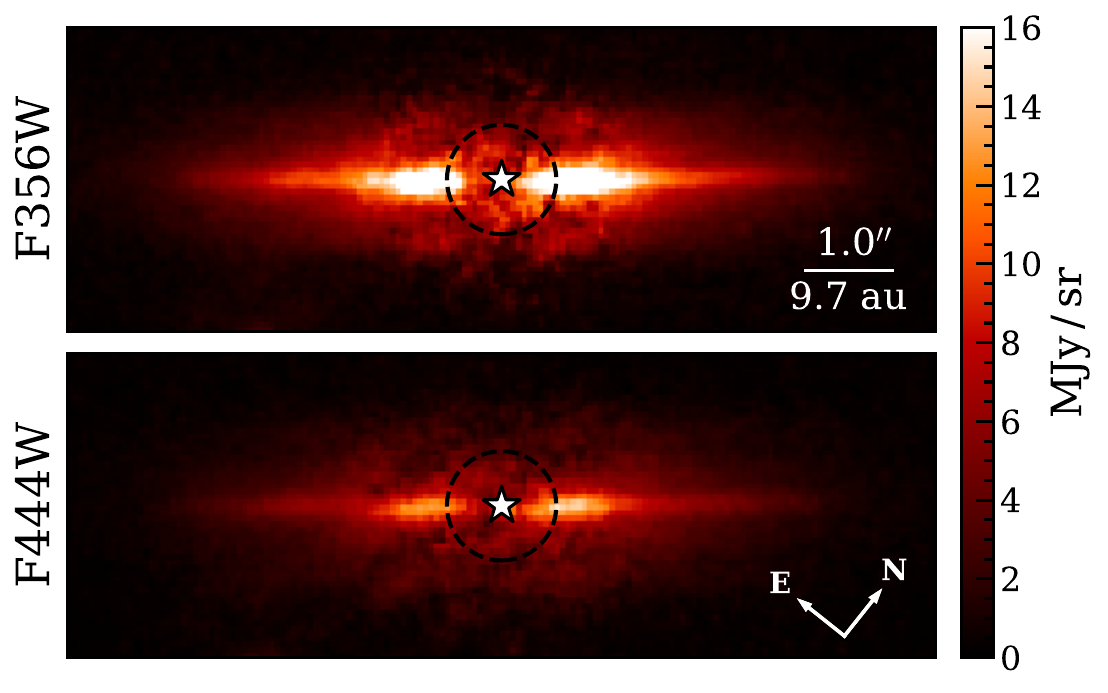}
\caption{Starlight-subtracted and roll-combined final JWST/NIRCam images of AU Mic in both F356W (top) and F444W (bottom). The images have been oriented such that the assumed disk major axis is parallel to the x-axis. To highlight the relative brightness of the disk at F356W, the images are displayed with the same linear color stretch. Both images have been smoothed with a $\sigma$ = 0.5 pixel gaussian for presentation, and are displayed within a $10\arcsec \times 3\farcs{}5$ \ac{fov}. The approximate coronagraphic inner working angle is indicated by the dashed black circle, while the approximate position of the occulted star is marked with a white star symbol. \label{fig:aumic_fig1}}
\end{figure*}

\begin{figure*}
\centering
\includegraphics[width=0.94\textwidth]{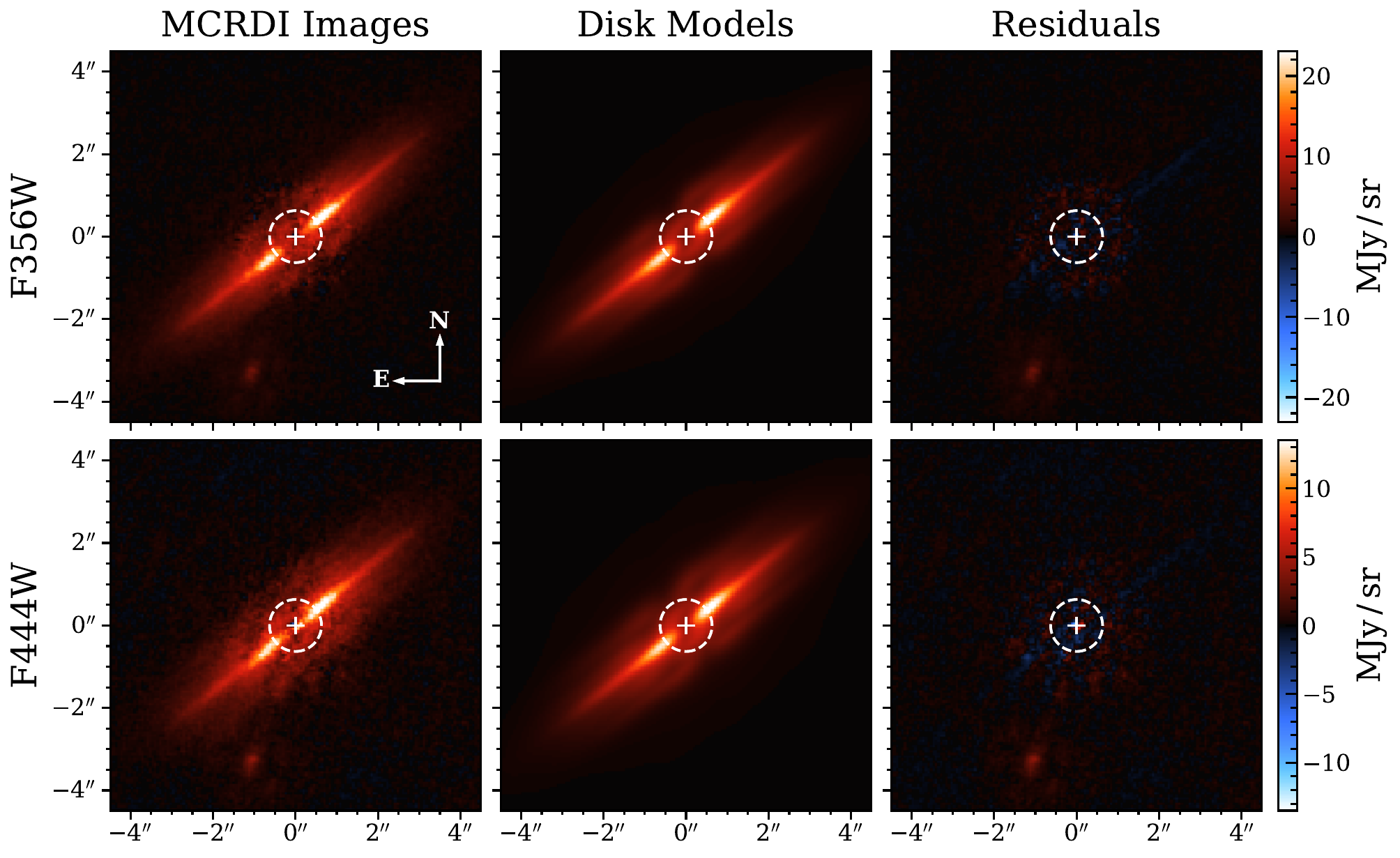}
\caption{For each filter (rows), the roll-averaged MCRDI final image (left), the best-fitting convolved model used as the constraint for MCRDI (middle), and the difference between the MCRDI result and the model image (right). The residuals use the same \edit1{linear} color stretch as the corresponding images to highlight that any differences are small compared to the total disk signal. A background source is visible in the lower left of the MCRDI images, at roughly $(-1\arcsec, -3\arcsec)$.\label{fig:mcrdi_models}}
\end{figure*}

\begin{figure*}
\centering
\includegraphics[width=0.94\textwidth]{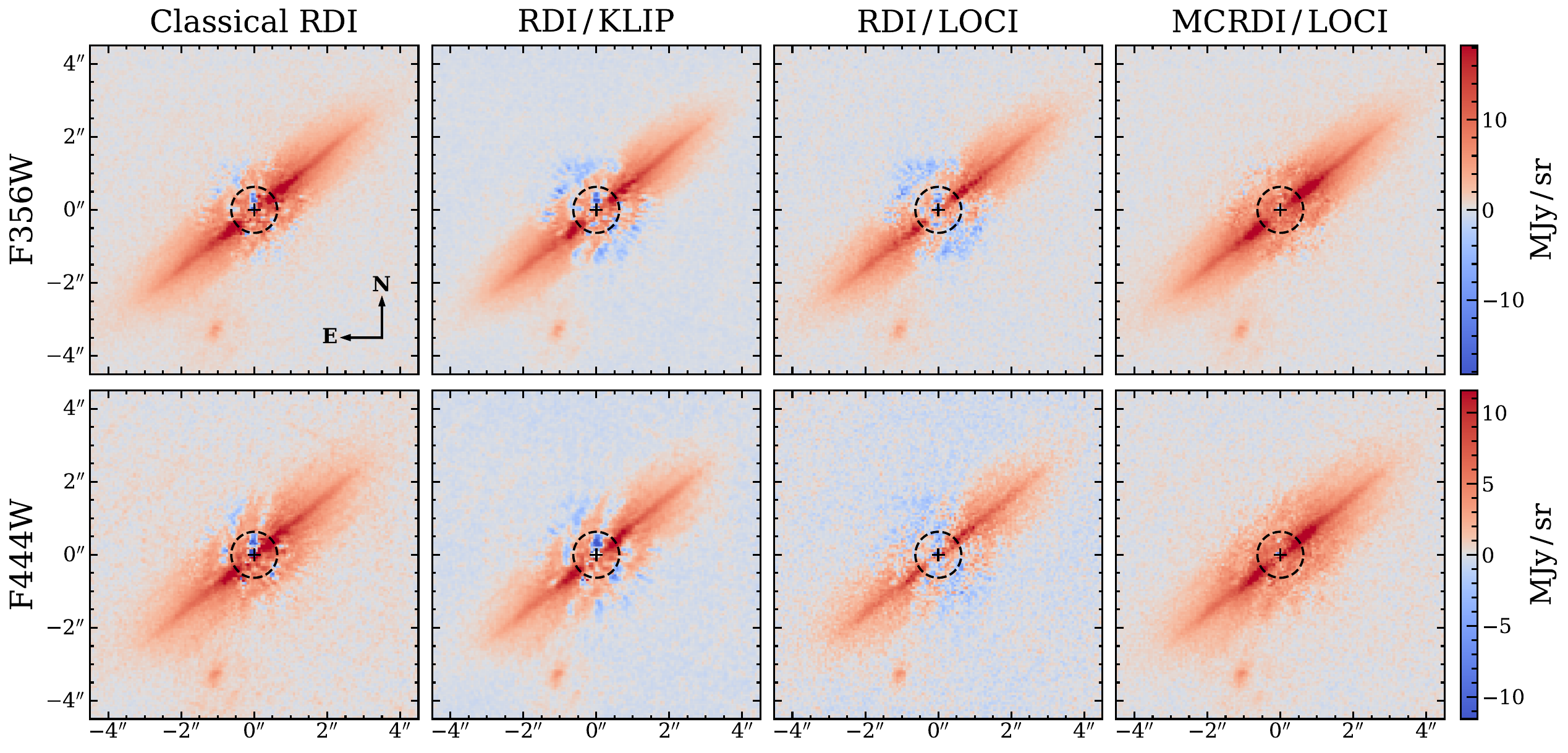}
\caption{Starlight-subtracted and roll-combined final images of AU Mic for the F356W (top row) and F444W (bottom row) filters using a variety of starlight subtraction techniques (as indicated by the column labels and described in \ref{sec:psfsub}). The black plus sign and dashed circle indicate the approximate coronagraph center and \ac{iwa} respectively. Interior to the \ac{iwa}, coronagraph transmission decreases and is at least partially responsible for the intensity deficit. The images for a given filter are displayed with the same linear color stretch, as indicated by the colorbar at the end of each row. A mapping symmetric about zero is chosen to highlight the effects of \ac{rdi} oversubtraction and speckle mismatch where present.\label{fig:aumic_reducs}}
\end{figure*}

The results of the MCRDI procedure are displayed in Figure \ref{fig:aumic_fig1}, while the MCRDI images, the corresponding model constraints, and the residuals are shown in Figure~\ref{fig:mcrdi_models}. The results for each of the four reductions in both filters are compared in Figure \ref{fig:aumic_reducs}.

\section{Detection of the AU Mic Debris Disk}
For both filters and all four starlight subtraction methods, the AU Mic debris disk is unambiguously recovered from the NIRCam observations (Figure \ref{fig:aumic_reducs}) — marking the first reported images of the disk at these wavelengths ($\sim$ 3–5$\,\micron$). The appearance of the disk in these data is generally consistent with the appearance at shorter wavelengths. Extended signal above and below the disk midplane (e.g., Figure \ref{fig:aumic_reducs}) are aliases of the disk produced by the off-axis ``lobes'' of the NIRCam diffraction pattern. \edit1{Modeling of AU Mic's spectral energy distribution in \citet{Matthews2015} found that the disk component is consistent with a 53K blackbody having a fractional luminosity of 3.5e-4 — for which emission would peak at $\sim$$55\,$$\micron$. Based on this, the observed light from the disk at the $3-5$$\,\micron$ can be assumed to be overwhelmingly composed of scattered starlight --- with thermal emission from the disk being negligible.}

Comparing the results from the different starlight subtraction methods, there are some notable differences. For both the RDI/KLIP and RDI/LOCI results, oversubtraction is evident from the predominantly negative background values and from the negative alias of the  coronagraphic stellar diffraction pattern at small separations ($r\lesssim2\arcsec$). Though the classical RDI procedure avoids these systematic oversubtraction effects, at separations $\lesssim 2\arcsec$ the results suffer from the non-optimized model of the diffracted starlight, which cannot account for changes to the pattern between the science and reference data (largely due to differences in coronagraph alignment). MCRDI provides the same high-fidelity as classical RDI at larger separations while permitting a much cleaner subtraction of the stellar diffraction pattern at small separations\footnote{See \citet{Lawson2022} for discussion of additional considerations that may be relevant for broader applications of the MCRDI technique.}.

We note that the differences between the KLIP and LOCI products are predominantly the result of the distinct optimization zones utilized (with KLIP using the full frame and LOCI using two narrow annular regions). A LOCI procedure utilizing the same regions as KLIP results in nearly indistinguishable results. 

The relative performance of classical RDI might be improved with the use of a dedicated reference sequence that adopts the small grid dither strategy (SGDS). Since the changes to the diffraction pattern between the science and reference sequences are generally dominated by differences in coronagraph alignment between the targets \citep[e.g.,][]{Girard2022}, observing the reference target at multiple dithers may increase the likelihood that the reference data includes a pointing that well-matches that of a given science pointing. However, as the dither offsets for NIRCam's SGDS are fixed \citep[and the dithering is precise;][]{Girard2022}, the value of this approach will be dictated by the accuracy of NIRCam's pointing; if the size of the dithers is significantly larger than the typical pointing error, it is unlikely that any of the dither offsets will provide an improved match for the science image. As \citet{Girard2022} note the possibility of further improvements for NIRCam target acquisition, the utility of the SGDS for typical classical RDI reductions is currently uncertain. Alternatively, SGDS reference data might be better leveraged by some misalignment-aware classical RDI procedure — in which a model of the starlight is constructed by combining multiple reference dithers to best emulate the misalignment of the science data.

\section{Companion Detection Limits}\label{sec:companion_limits}
\begin{figure*}
\centering
\includegraphics[width=0.95\textwidth]{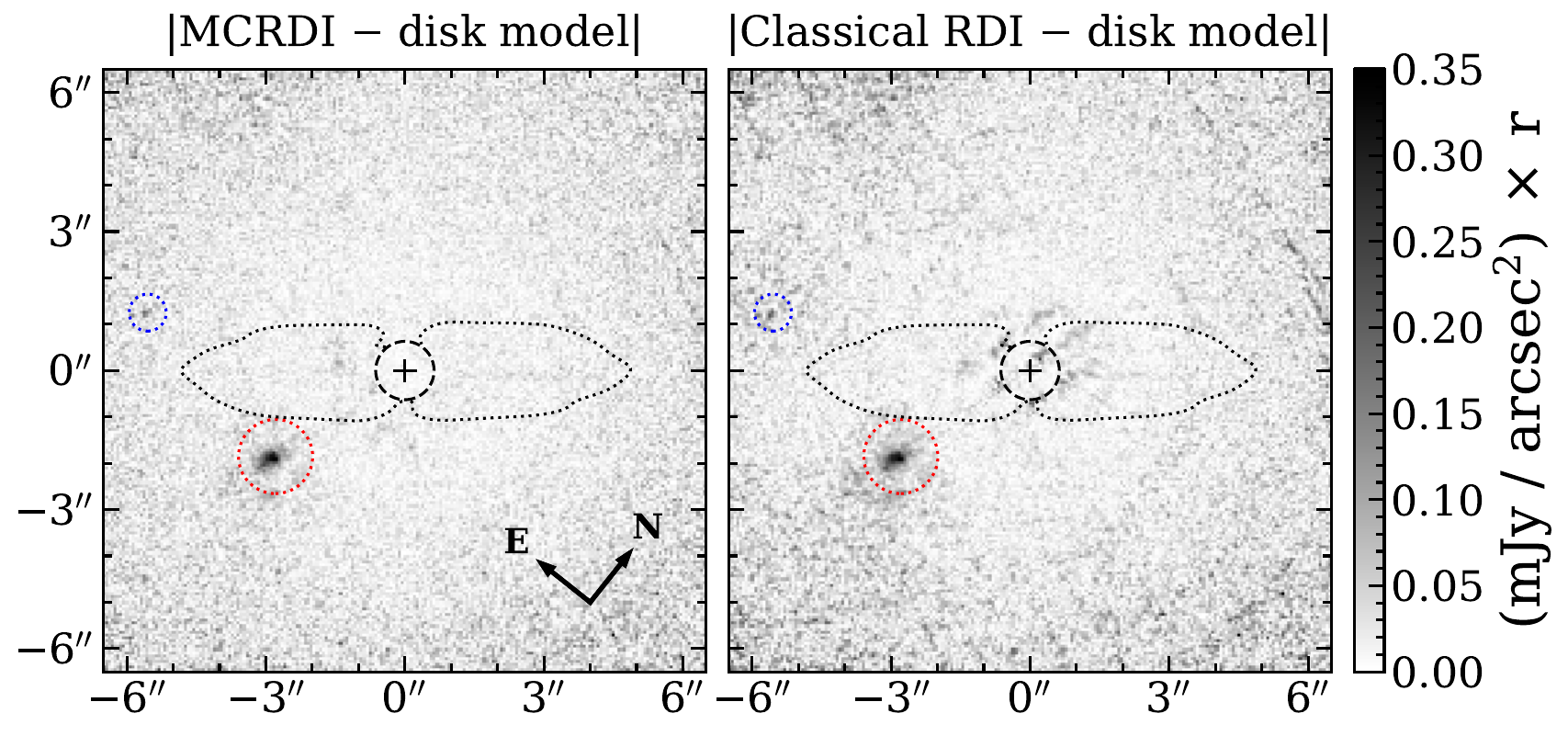}
\caption{The absolute value of disk-model-subtracted images for MCRDI and classical RDI reductions in F444W, oriented as in Figure \ref{fig:aumic_fig1}. Each image has been multiplied by the projected \edit1{radial} separation in units of arcsec \edit1{and is presented with a linear color stretch}. The dotted black contour outlines the extent of the disk model, with two previously identified background sources indicated by the blue and red dotted circles. As a reference for the depth of the imagery: based on \citet{Linder2019} models, the brightness of the background source circled in red would correspond to a foreground companion with a mass comparable to Saturn's, while the brightness of the source circled in blue would correspond to a mass of $\sim 0.12\;M_J$. \label{fig:companion_search}}
\end{figure*}

Inspecting the final images and disk-model-subtracted images, we identify no candidate companions near the plane of the disk. For reference, disk-model-subtracted images for MCRDI and classical RDI reductions of the F444W observations are shown in Figure \ref{fig:companion_search}. A number of point-sources appear within the field of view, but are consistent with previously observed background sources based on AU Mic's proper motion \citep[assuming Gaia DR3 values of $\mu_\alpha = 281$ mas/yr and $\mu_\delta = -360$ mas/yr;][]{GaiaDR32022}. For example, adjusting the locations of the two off-axis sources appearing in Figure \ref{fig:companion_search} for proper motion, the results fall within $\sim 50$ mas of the locations of sources appearing in 2017 HST/STIS coronagraphy of AU Mic \citep[originally published in][]{Wisniewski2019,Grady2020}. With no plausible companions detected, we proceed with analysis of the sensitivity of the data to assess both the companions we would likely have detected and what companions might still remain. 

Computation of conventional contrast curves is complicated in this case by the fact that the flux of the occulted target star, AU Mic, cannot be directly measured from these data. While a synthethic stellar spectrum could be adopted for AU Mic's photosphere or otherwise used to convert existing photometry (e.g., for similar WISE filters) to NIRCam photometry, this nevertheless introduces an additional layer of model dependence. In light of this, we focus on the more direct calculation of companion detection limits in terms of companion brightness.

In this process, we determine limiting companion fluxes as a function of projected separation using forward-modeling. Since companions are generally expected to orbit within or near the plane of the disk, we exclusively consider coplanar companions here (assuming a position angle of $128\fdg48$ and an inclination of $90\degr$ for simplicity). For an array of projected separations spanning approximately $0\farcs1$ to $12\arcsec$ along both the southeastern and northwestern extents of the disk, we compute the corresponding position for each roll of the AU Mic data. Then, we use the \texttt{WebbPSF\_ext} package to generate a \ac{psf} at each position using the closest \ac{opd} map and configured to match the observing configuration of the data. Each synthetic companion image is normalized to have a total flux of 1 mJy times the coronagraph transmission at the companion's location (i.e., normalized to correspond to a source whose unocculted flux is 1 mJy). For each roll angle, our procedure considers the fit coronagraph offset from Section \ref{sec:imreg} when generating the model images. Subsequently, we simulate the effects of \ac{rdi} on these sequences — as typical for RDI/LOCI forward modeling \citep[e.g.,][]{Currie2019} — ultimately producing an oversubtracted and roll-combined model image. Additionally, we consider a scenario in which MCRDI is used to improve throughput (primarily at small separations, where RDI oversubtraction more significantly attenuates companion flux). For simplicity, we assume that MCRDI is able to effectively suppress all oversubtraction introduced by the companion, such that the coronagraph is the only significant limitation on companion throughput.

To determine the $1\sigma$ noise level in our data, we generate radial noise profiles for the MCRDI reductions of both filters. For this, we first replace the value of each pixel in the image with the sum of the values within a FWHM-diameter aperture (2.29 and 2.84 pixels for F356W and F444W, respectively). Then, for each radial separation from the star, $r_i$, the corresponding noise level is taken to be the finite-element-corrected standard deviation \citep{Mawet2014} of all pixels with stellocentric separations, $r$, falling in $(r_i - 0.5 \cdot \rm{FWHM})< r < (r_i + 0.5 \cdot \rm{FWHM})$. To mitigate the impact of the disk, which would otherwise serve to artificially inflate the noise levels, we subtract the best-fitting MCRDI disk model (see Section \ref{app:mcrdi_models}) before measuring the noise. Additionally, we exclude the vicinity of five background sources from this calculation. The \ac{snre} map for the forward-modeled synthetic companion image is then taken to be the aperture-summed companion image divided by the noise map computed for the data. Since the maximum \ac{snre} achieved across a companion's \ac{psf} may not be coincident with the injected candidate position — because the \ac{psf} spans a range of radial separations (and thus a range of noise levels) — we do not simply adopt the value at the injected position as the throughput \ac{snre}. Instead, we adopt the largest \ac{snre} value within 12 times the FWHM of the candidate position\footnote{This radius is chosen to include the six bright off-axis ``nodes" of the NIRCam \ac{psf}.}. This value is the throughput \ac{snre} for a companion of 1 mJy at the prescribed location. Since the forward-modeled flux for a companion model scales linearly with the input flux, the 5$\sigma$ limiting flux in mJy is taken to be 5 divided by the peak throughput \ac{snre} of the 1 mJy model. When computed this way, the limits along the northwestern and southeastern extents of the disk are very similar (with small differences resulting from the asymmetry of the PSF and the misalignment of the coronagraph). Thus, for simplicity, we average the two values for a given separation when presenting sensitivity curves.

We note that the procedure outlined above does not account for residual disk flux resulting from inaccuracy of our disk model. Though these residuals affect our noise calculation (increasing our measured noise), they are not considered in the measurement of the ``signal" component of the forward-modeled \ac{snre}. In practice, where our disk model is brighter than the true underlying disk in the data (negative residuals), the flux of any coincident companions would appear to be diminished — and vice versa. Counter-intuitively, this could result in a companion below the limits we report apparently manifesting with sufficient \ac{snre} to be `recovered'. Given the faintness of the disk residuals (see Figures \ref{fig:companion_search}, \ref{fig:mcrdi_models}), we make no effort to account for this contribution quantitatively in this work. 

The use of the MCRDI reduction for measuring the noise profile means that the detection limits achieved for ``RDI" are representative of a reduction using MCRDI to mitigate oversubtraction from the disk, but making no effort to suppress oversubtraction for the considered off-axis companion. For a LOCI or KLIP reduction that does not account for the presence of the disk, simply forward modeling point-sources would not be appropriate for assessing the faintest recoverable companions. If not suppressed, the disk will serve to significantly reduce the throughput circumstellar flux throughout the \ac{fov} — including that of any companions \citep[e.g.,][]{Lawson2022}. To accurately account for the effects of starlight subtraction using any technique in which the stellar model is optimized by comparison with the data itself or in which the target data is used to build the \ac{psf} model (e.g., any common technique besides classical RDI), the entire circumstellar scene must be considered holistically in forward modeling. To address this, an ``injection-recovery" approach \citep[e.g.,][]{Carter2022} could be used in some scenarios. However, for an edge-on disk system like AU Mic, any position at which a companion would likely manifest is coincident with a non-negligible quantity of disk flux. So, while injection-recovery would account for the effects of the rest of the circumstellar scene on oversubtraction (and self-subtraction in the case of angular differential imaging), we would still require some means of disentangling companion and disk signal. At some point, a model of the disk must be assumed in order to assess the sensitivity of these data to companions.

To map fluxes in the JWST/NIRCam filters to companion masses, we use the \texttt{species} Python package \citep{Stolker2020}. In \texttt{species}, companion properties (luminosity, surface gravity, effective temperature, etc.) are interpolated from grids of synthethic isochrones for a given age (assuming an age of 24 Myr for AU Mic) and companion mass. Synthetic spectra are then interpolated to produce a spectrum corresponding to these companion properties. Finally, \texttt{species} extracts photometry from the resulting spectrum. For masses of 0.6 $M_{\rm J}$ and above, we use the AMES-Cond isochrones from \citet{Allard2001}. For lower mass objects, none of the isochrone sets built into \texttt{species} have coverage at the age of AU Mic. Instead, we manually introduce the cloud-free petitCODE isochrones for low-mass objects provided by \citet{Linder2019} for masses of 0.5 $M_{\rm J}$ and below. In either case, the isochrone properties are paired with AMES-Cond synthetic spectra \citep{Allard2001} to estimate would-be companions' photometry. This ultimately provides valid synthethic photometry for masses as low as $\sim$ 0.03 $M_{\rm J}$. Comparing the photometry from \texttt{species} using \citet{Linder2019} isochrones and AMES-Cond spectra with the precomputed photometry for the \citet{Linder2019} isochrones shows F444W fluxes that are comparable, but F356W fluxes that are much fainter for the precomputed photometry. As such, the companion detection limits for F444W are likely the more robust of the two filters. As the F444W filter is the more sensitive of the two to low-mass companions, this does not directly impact the faintest recoverable companions.

\begin{figure*}
    \centering
    \includegraphics[width=0.75\textwidth]{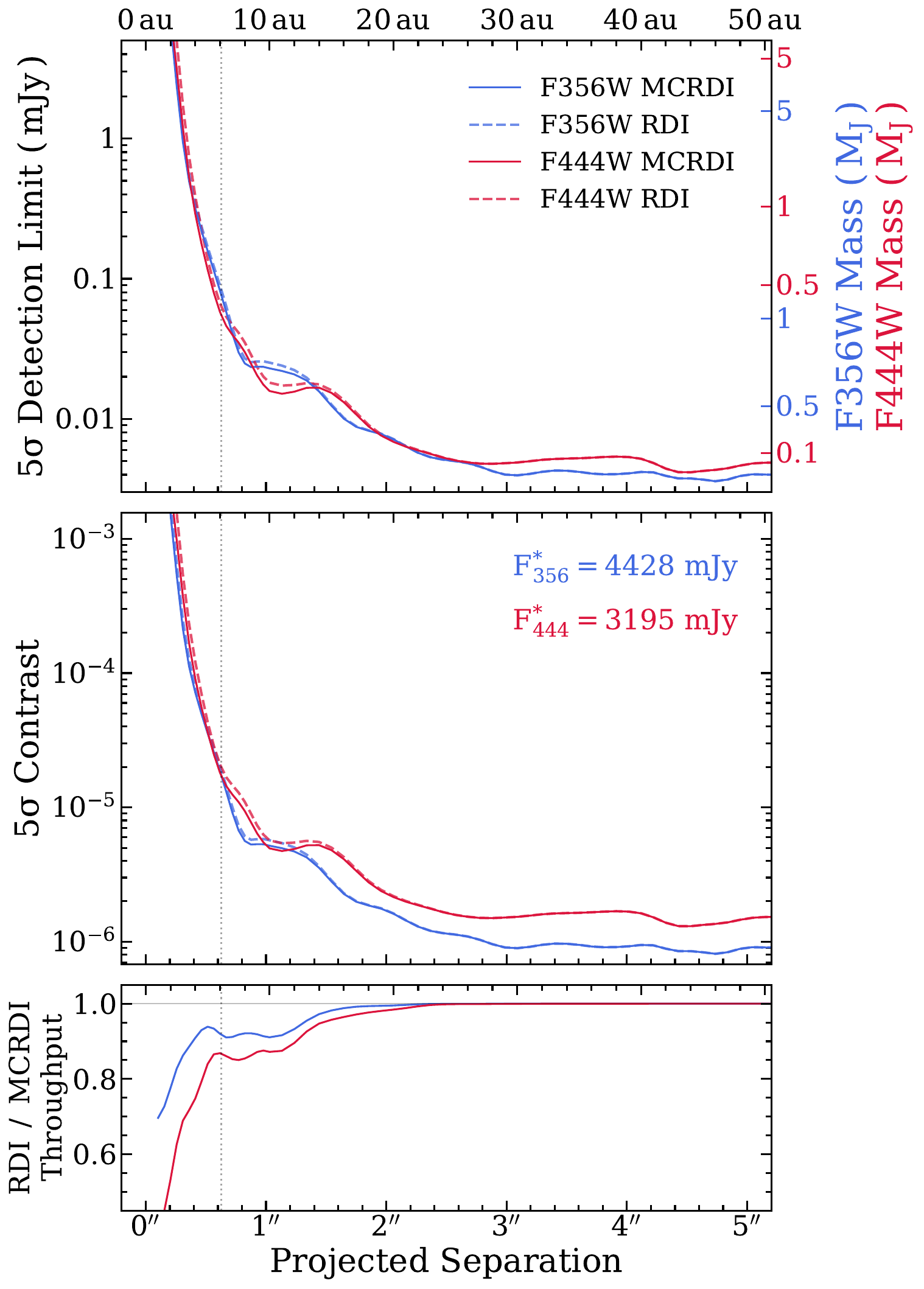}
    \caption{\textbf{Top}: 5$\sigma$ companion detection limits in mJy for both filters — computed as described in Section \ref{sec:companion_limits}. The right y-axis provides the corresponding mass limits for each filter in Jupiter masses — using AMES-Cond and cloud-free petitCODE isochrones as outlined in Section \ref{sec:companion_limits}. \textbf{Middle}: $5\sigma$ contrast curves for the same reductions, adopting the model-based stellar fluxes for AU Mic indicated in the upper right corner (see Section \ref{sec:companion_limits}).
    \textbf{Bottom}: the ratio of the companion throughput for the RDI-LOCI and MCRDI procedures. The improvement for MCRDI persists up to approximately the edge of the outer optimization region used for construction of the starlight model ($r = 2\farcs205 = 35$ pixels). For all panels, the x-axis indicates the projected stellocentric separation of the measurement — in arcsec on the lower x-axis and in au on the upper x-axis. \edit1{The relatively gentle slope of the coronagraph's transmission profile permits some sensitivity to sources at smaller separations than the \ac{iwa} of 0\farcs63 (dotted gray line).}
    \label{fig:detection_limits}}
\end{figure*}

To convert flux limits to approximate contrasts, we scale a synthetic photosphere approximating AU Mic's mass (0.5 $M_\odot$) and age (24 Myr) to match 2MASS J- and H-band photometry of AU Mic \citep{Skrutskie2006}, and then extract photometry from the resulting spectrum. This yields approximate F356W and F444W fluxes of 4428 mJy and 3195 mJy respectively. Using photometry from the F335M target acquisition images \citep[multiplied by a factor of 516 for use of the neutral density square; \texttt{WebbPSF\_ext}, ][]{Leisenring2021} to rescale the same synthetic model (in place of 2MASS photometry) yields very similar F356W and F444W fluxes of 4447 mJy and 3209 mJy respectively. The $5\sigma$ flux, mass, and contrast limits are presented in Figure \ref{fig:detection_limits}.

\begin{figure}
\centering
\includegraphics[width=0.47\textwidth]{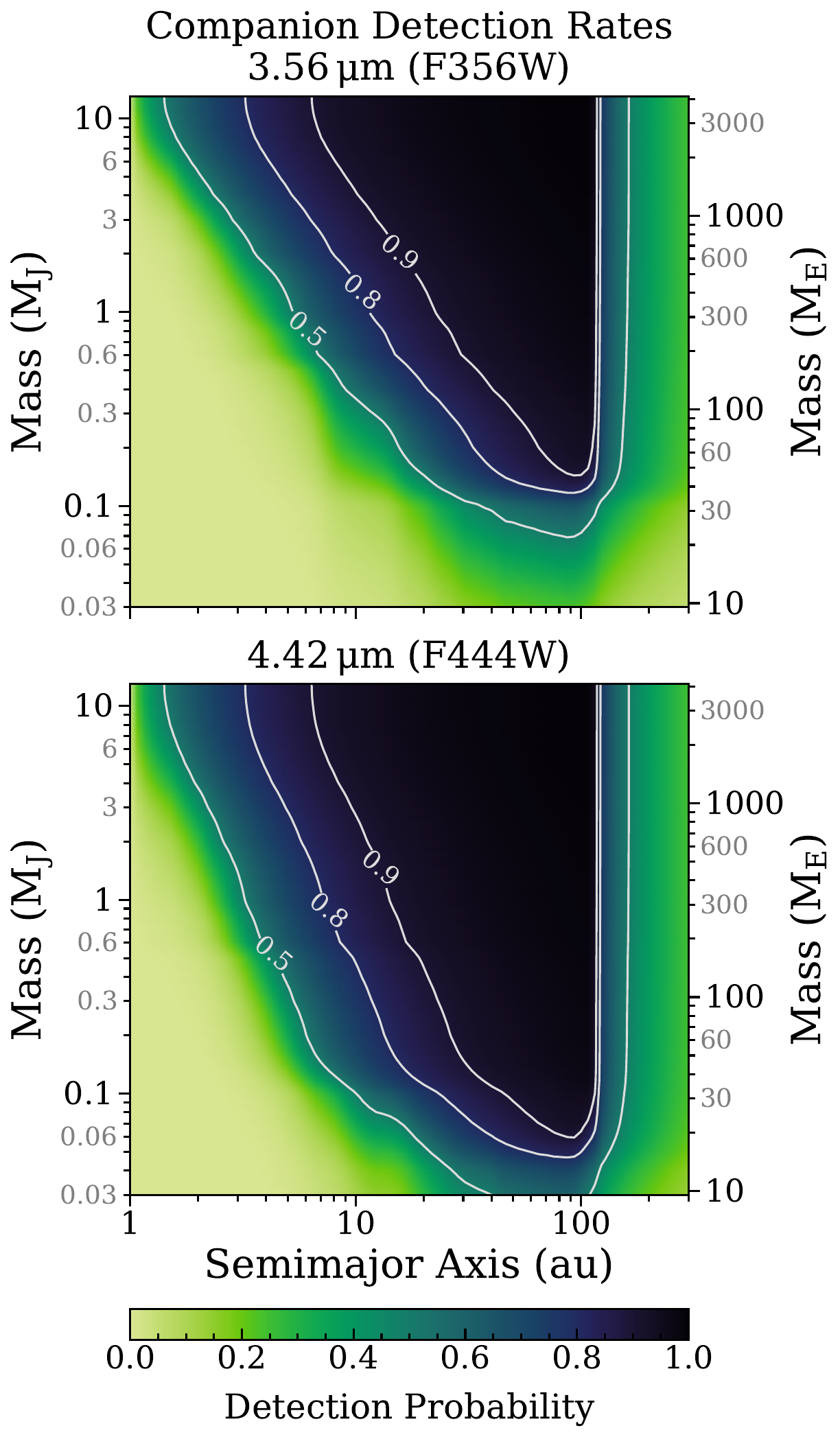}
\caption{Maps of orbit-averaged companion detection probabilities as a function of companion semimajor axis (au) and companion mass (Jupiter masses on the left axis, and Earth masses on the right). For the F444W observations, the map indicates detection probabilities for analogs of Jupiter, Saturn, and Neptune (i.e., matching semimajor axis and mass) of 0.73, 0.76, 0.70 respectively.\label{fig:tongue_plots}}
\end{figure}

We then compute maps of companion detection rates (``tongue plots") for the MCRDI reductions using this information and assuming companions following circular orbits that are coplanar to the disk — which is assumed to be optically thin. For each synthetic companion mass, we consider a logarithmic grid of semimajor axes with 300 values from 1 au to 1000 au. For each semimajor axis and mass, we generate a sample every $0\fdg01$ over a full orbit, determine the projected separation for each sample, and then linearly interpolate the \ac{snre} for that sample from the previously computed $5\sigma$ detection limits. This \ac{snre} is then mapped to a detection probability assuming a normal distribution (e.g., that a $3\sigma$ detection corresponds to a 99.7\% chance of detection). The overall detection rate for a companion at a given separation and mass is taken to be the average of the 36000 samples over the full orbit. The companion detection maps for both filters are shown in Figure \ref{fig:tongue_plots}. We emphasize that these are not presented in terms of projected separations, but rather the true semimajor axes for companions on circular orbits.

\subsection{Validation with Companion Injection}\label{sec:companion_injection}
As a proof of concept for these detection limits, we injected two low-mass companions into the unsubtracted Stage 2 F444W products for AU Mic: a $0.1 \; M_J$ planet at a projected separation of $2\farcs71$ along the southeastern extent of the disk, and a $0.2 \; M_J$ planet at a projected separation of $1\farcs72$ along the northwestern extent of the disk. For this purpose, we use PSFs from \texttt{WebbPSF\_ext} and again assume fluxes based on \citet{Linder2019} isochrones with AMES-Cond spectra (via \texttt{species}). The positions were drawn from circular coplanar orbits with semimajor axes of 30 and 20 au respectively in order to place the companions at projected separations where they would be slightly brighter than the detection limits of Figure \ref{fig:detection_limits}. 

\begin{figure*}
\centering
    \includegraphics[width=0.95\textwidth]{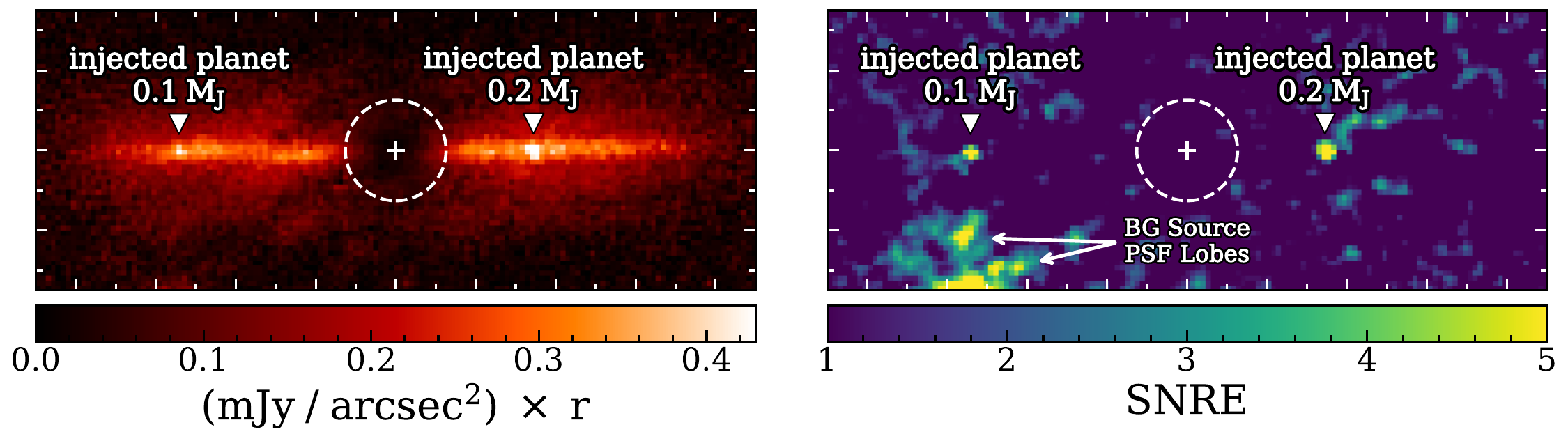}
\caption{Final F444W MCRDI products following the injection of two low-mass companions into observations of AU Mic (see Section \ref{sec:companion_injection}). The images are oriented such that the disk major axis is aligned with the horizontal axis. Left: the final MCRDI image, which has been multiplied by the projected \edit1{radial} separation in units of arcsec to improve the visualization. Right: the SNRE map for the residuals between the MCRDI image and the MCRDI disk model. For both panels, the location of each injected companion is indicated by a white caret marker and annotation. Major and minor axis ticks delineate separation in increments of $1\arcsec$ and $0\farcs5$ respectively. \edit1{A linear color stretch is used for both images.}\label{fig:planet_injection}}
\end{figure*}

We then reran the full MCRDI disk model optimization procedure exactly as for the real data to verify that these companions would be recoverable in the resulting disk-model-subtracted results. During MCRDI optimization, we made no effort to mask the companions — meaning that the resulting disk model slightly over-estimates the true disk's brightness (by $\sim 2\%$) to compensate for the companion flux. As both companions are visible in an initial ``unconstrained" RDI/LOCI reduction, the final result could be improved by masking their locations throughout the procedure. Rerunning the procedure while masking the region within 5 times the FWHM of each companion had no perceptible impact on the final MCRDI image ($<0.1\%$ average change in brightness; the over-bright disk in the unmasked scenario compensates for the companions' flux in the least-squares optimization of the stellar PSF model), but does mitigate the over-brightening of the disk model. As such, there is a small difference in the brightness of the companions when the disk model is subtracted from the final MCRDI image in each scenario. For this demonstration, we proceed with the simpler unmasked version. Computing \ac{snre} as before, the $0.1 \; M_J$ and $0.2 \; M_J$ companions have peak \ac{snre} values of \edit1{6.6} and \edit1{9.5} respectively — manifesting with $> 5 \sigma$ significance, as anticipated. The resulting MCRDI image and the disk-model-subtracted \ac{snre} map are shown in Figure \ref{fig:planet_injection}.

When applying MCRDI to data with visible companion candidates in nominal reductions, point sources could also be explicitly added to the MCRDI model instead. Though this increases the complexity of the optimization problem compared to simply masking them, it will likely improve the result when a candidate lies well within the speckle-limited regime (i.e., near the \ac{iwa}), and has the benefit of providing model-based measurements of companion fluxes as a byproduct. In application to data with companions that are much brighter relative to the disk: if not somehow addressed, the effect of the companions on the disk model will be more significant. In such a scenario, either masking or explicitly including the companions in the MCRDI \ac{css} model is recommended. 

\subsection{Discussion of Companion Detection Limits}
The limiting companion flux of $\sim 0.004$ mJy at larger separations in Figure \ref{fig:detection_limits} corresponds to $5\sigma$ flux contrasts of approximately $9.0\times10^{-7}$ and $1.3\times10^{-6}$ for F356W and F444W respectively. These contrasts are consistent with the contrasts reported in, e.g., \citet{Carter2022} and \citet{Girard2022}. In terms of companion masses, these contrasts permit exceptionally deep limits as a result of the relatively low luminosity of the parent star — with F444W showing sensitivity to $<0.1 \; {\rm M_J}$ companions. The detection maps computed using forward modeling (Figure \ref{fig:tongue_plots}) show, for example, that for the F444W filter, our observations have a $\sim 92\%$ chance of detecting any Saturn mass ($\sim 0.3 \; M_{J}$) companion with a semimajor axis of 25 au (falling just inside the disk's fiducial radius). 

Based on modeling of the AU Mic disk in ALMA dust continuum emission, \citet{Pearce2022} and \citet{Vizgan2022} each provide estimates for a yet-unseen companion that might be responsible for shaping the inner edge of the disk. \citet{Pearce2022} find a companion mass and semimajor axis of $\sim 0.44 \; M_{J}$ and $\sim$ 21.9 au (from an inner disk radius of 28.7 au), while \citet{Vizgan2022} find a companion mass and semimajor axis of $\sim 0.34 \; M_{J}$ and $\sim$ 17 au (from an inner disk radius of 22.1 au). Permitting the aforementioned assumptions, our analysis indicates that we would have overall detection probabilities of $\sim 91\%$ and $\sim 87\%$ (respectively) for these planets in F444W. If such a planet remains, it would most likely be at very small angular separations at the epoch of these data — as each of the proposed objects would manifest with $\ac{snre}\geq 5$ until $r \lesssim 0\farcs6$. 

\citet{Sezestre2017} conducted a dynamical analysis of a possible unseen ``parent body" responsible for the fast-moving structures reported in \citet{Boccaletti2015} — exploring models for both an orbiting parent body (e.g., a planet) and a static parent body (e.g., a dust cloud resulting from a large planetesimal collision). The best-fit solution in the case of the orbiting parent body scenario would manifest at a projected separation of $\sim 4.6$ au ($0\farcs47$) to the northwest at the epoch of the NIRCam data presented here (see right panel of Figure 8 in \citealt{Sezestre2017}). For F444W, a companion at this location would have been detected at $\geq 5 \sigma$ for masses $\gtrsim 0.54 \; \rm M_J$ --- placing a strong upper limit on the mass of the parent body in this scenario. Overall, future studies combining similar analysis as \citet{Sezestre2017} with the detection limits provided herein should enable a much clearer understanding of the source of the fast moving features observed in AU Mic's disk.

\edit1{Overall, the non-detection of outer giant planets and these deep detection limits suggest the likely absence of outer giant planets in the AU Mic system. This echoes the results of \citet{Daley2019}, who analyzed the vertical dust distribution of AU Mic's disk and ultimately inferred that its thin vertical size should preclude the presence of perturbers more massive than $\sim\,$$\rm 2 \; M_\earth$ in the outer disk. Given the lack of outer giant planets, which would otherwise prevent volatiles from streaming inward \citep[e.g.,][]{Clement2022}, transmission spectra for the inner planets should be relatively likely to reveal atmospheres that are rich in water or other volatiles.}

\section{Disk Surface Brightness and Color}\label{sec:sb}
Here, we provide measurements of \ac{sb} and color for AU Mic's disk from reductions using classical RDI, RDI/KLIP, and MCRDI. A subsequent study focusing on analysis of the disk in these data will further investigate these measurements and their implications.

We measure \ac{sb} within circular apertures with radii of 4 pixels ($0\farcs252$) placed along the spine of the disk. For simplicity, the spine positions are taken to be the analytic spine for a ring-like forward-scattering disk with a fiducial radius of 35 au, an inclination of 89\degr, and a position angle of $128\fdg48$  (i.e., falling along an ellipse for $r_{proj} < 35$ au, and along the major axis otherwise). These measurements are made using the derotated and roll-averaged final image products\footnote{The results showed no statistically significant differences when instead averaging surface brightness measurements from individual rolls without derotation.}.

Uncertainties on these measurements are estimated by making a set of like-measurements at the same projected separation but differing position angles (in increments of $3\degr$ in the range [0\degr, 357\degr], for 119 total sets), and then calculating the median absolute deviation among the values at each position. To mitigate the effect of the disk itself on the uncertainty estimates, the disk model used with MCRDI is subtracted from each image before the measurements are made (in the case of the RDI/KLIP reductions, the disk model is first forward-modeled to induce appropriate oversubtraction). Generally, the resulting uncertainties are much larger than the differences between the MCRDI and classical RDI measurements beyond the small separations where classical RDI suffers from stellar residuals. Meanwhile, the RDI/KLIP measurements are affected by oversubtraction — which is not considered in these uncertainties — leading to measurements that are significantly discrepant with those of the other reductions.

\begin{figure*}
    \centering
    \includegraphics[width=0.98\textwidth]{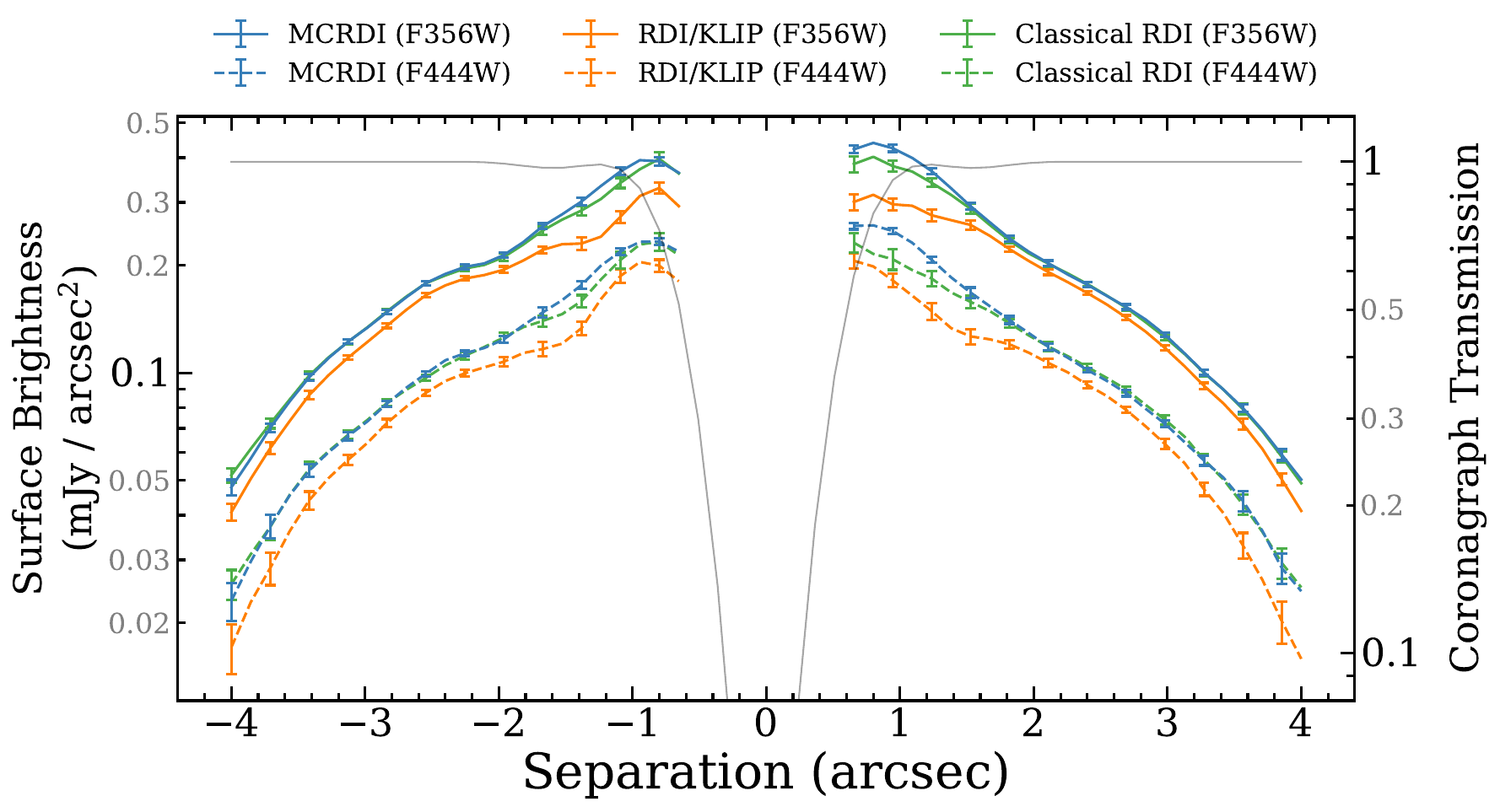}
    \caption{Surface brightness measurements along the spine of AU Mic's disk in F356W (solid lines) and F444W (dashed lines) measured in circular apertures ($r = $ 4 pixels $=0\farcs252$) for images produced by three distinct starlight subtraction techniques: MCRDI (blue), RDI/KLIP (orange), and classical RDI (green). The x-axis indicates the stellocentric separation projected onto the assumed disk major axis, with negative values corresponding to the southeastern extent of the disk. The roll-averaged coronagraph transmission profile is displayed as a light gray line (measured at the same locations and using the same aperture size), with values given along the right y-axis. To improve readability, error bars are plotted for alternating surface brightness measurements. \label{fig:sb}}
    \end{figure*}

\begin{figure*}
    \centering
    \includegraphics[width=0.92\textwidth]{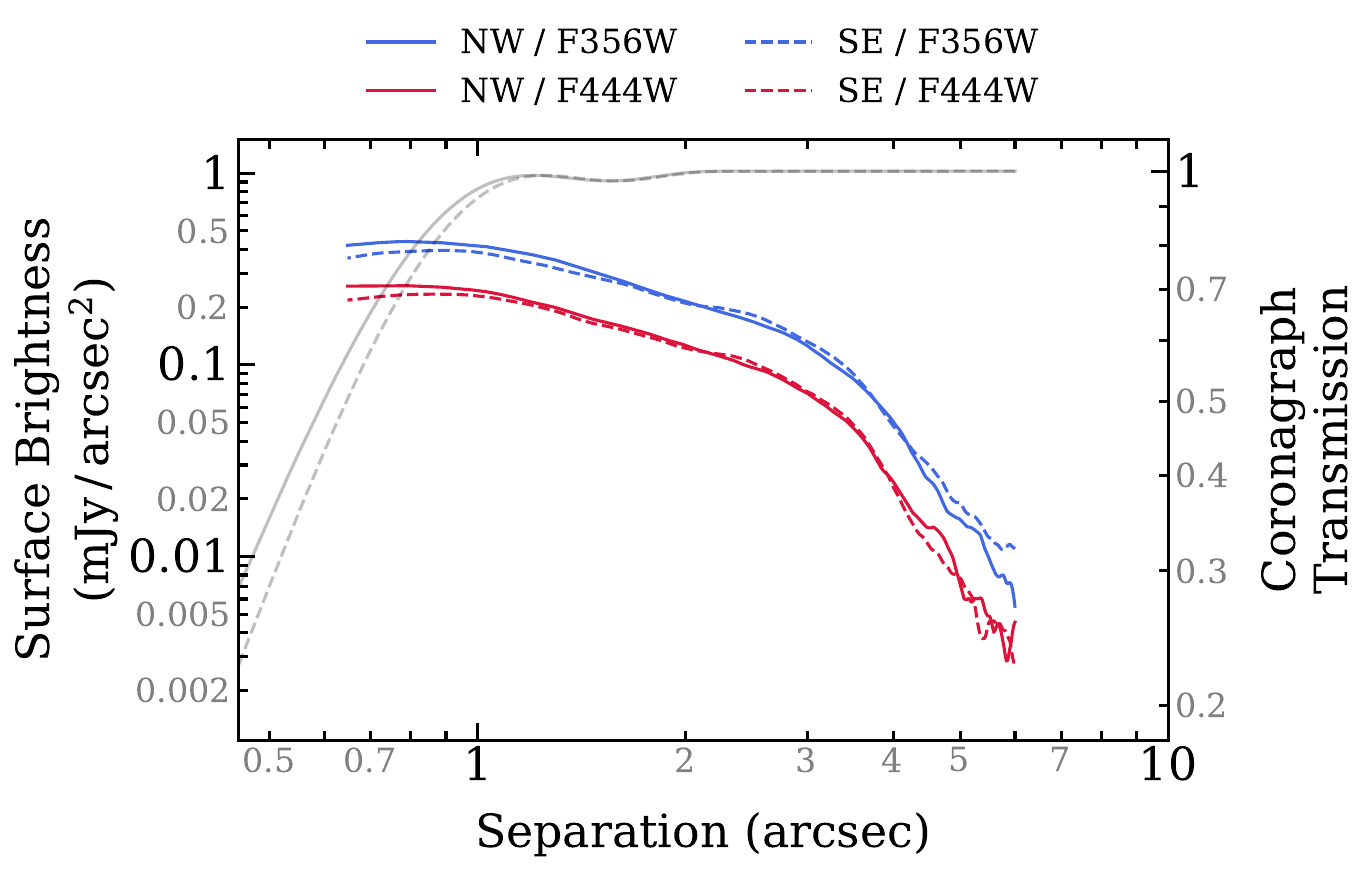}
    \caption{\edit1{Surface brightness profiles for the MCRDI reductions (F356W in blue and F444W in red) presented in a log-log scale and including separations to $6\arcsec$. The roll-averaged coronagraph transmission profiles are displayed in light gray, with values given along the right y-axis. For both surface brightness and coronagraph transmission, the northwestern (NW) and southeastern (SE) profiles are indicated by solid and dashed lines (respectively). Error bars are excluded to improve readability. \label{fig:log_log_sb}}}
    \end{figure*}

\begin{figure*}
    \centering
    \includegraphics[width=0.95\textwidth]{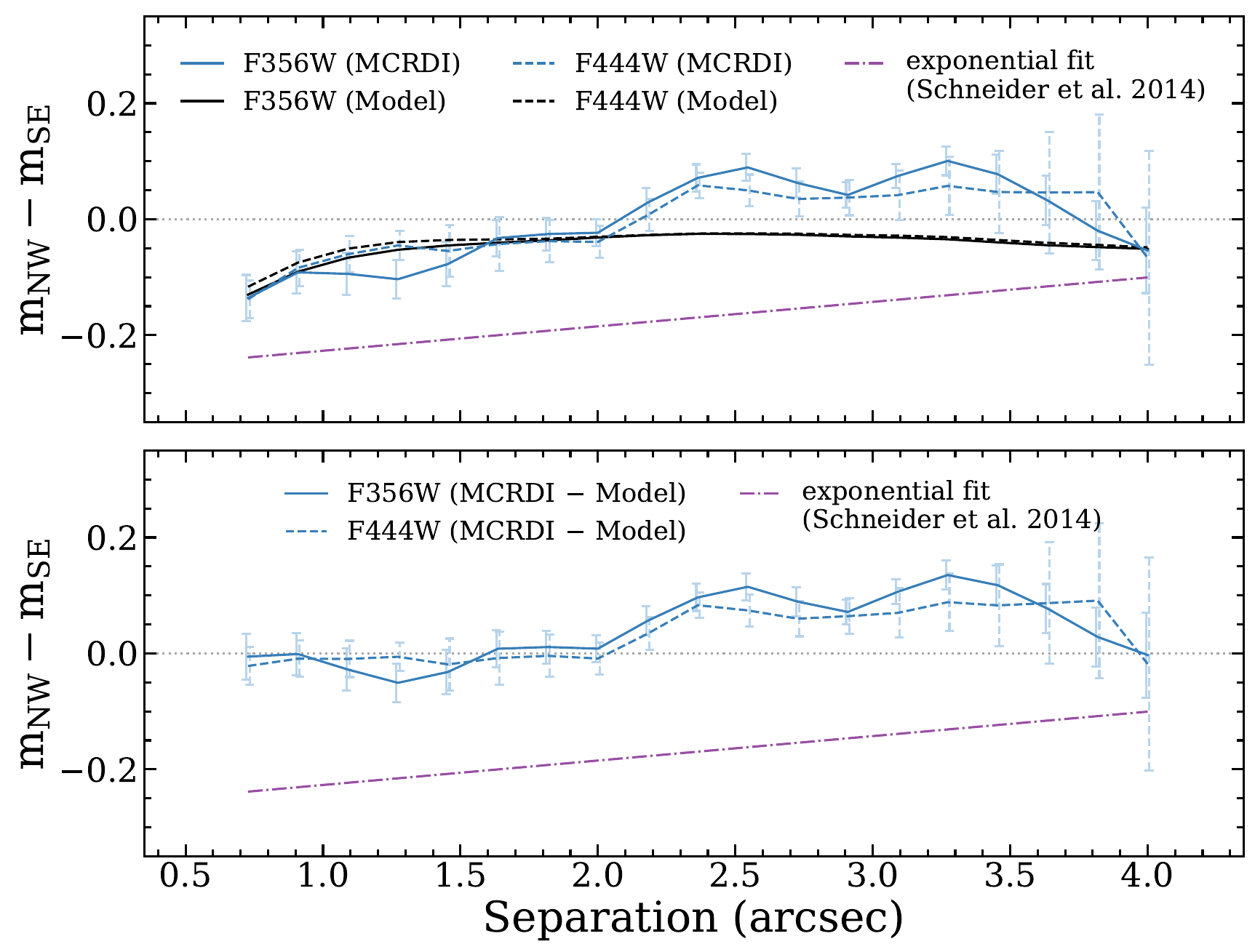}
    \caption{\edit1{A comparison of the surface brightness profiles for the northwestern (NW) and southeastern (SE) extents of the disk for the MCRDI reductions (blue) in magnitudes. In the upper panel, profiles for the convolved disk models (black), whose unconvolved images are perfectly symmetric, are included to highlight effects artificially induced by coronagraph misalignment and by the asymmetry of the PSF itself. In the lower panel, the artificial asymmetry has been corrected by subtracting the model asymmetry measurements of the upper panel from those of the MCRDI measurements. For context, the dash-dotted purple line shows the expected asymmetry based on the exponential fits to HST/STIS surface brightness profiles for AU Mic's disk from \citet{Schneider2014}. In both panels, error bars are offset slightly in the x-axis direction to improve readability.}
    \label{fig:sb_asymmetry}}
    \end{figure*}

\Ac{sb} profiles for both filters are shown in Figure \ref{fig:sb}. For the MCRDI reductions, we also provide \edit1{a) a log-log plot of surface brightness in Figure \ref{fig:log_log_sb}, and b) a comparison of the brightness between the southeastern (SE) and northwestern (NW) extents of the disk in Figure \ref{fig:sb_asymmetry}}.

To measure the disk's F356W versus F444W color, we adopt the model-based stellar fluxes used to estimate contrast in Section \ref{sec:companion_limits}: $F^*_{356} = 4428$ mJy and $F^*_{444} = 3195$ mJy. Denoting a measurement of the disk \ac{sb} with superscript $d$, the disk color in magnitudes is computed as the difference between the \ac{sb} color and the stellar color: 

\begin{equation}
    \begin{split}
    \Delta(m_{356} - m_{444}) = -2.5 \left[ \log_{10} \frac{F^d_{356}}{F^d_{444}} - \log_{10} \frac{F^*_{356}}{F^*_{444}} \right] \\
    = -2.5 \log_{10} \frac{F^d_{356} F^*_{444}}{F^d_{444} F^*_{356}}
    \end{split}
\end{equation}

\begin{figure*}
\centering
\includegraphics[width=0.95\textwidth]{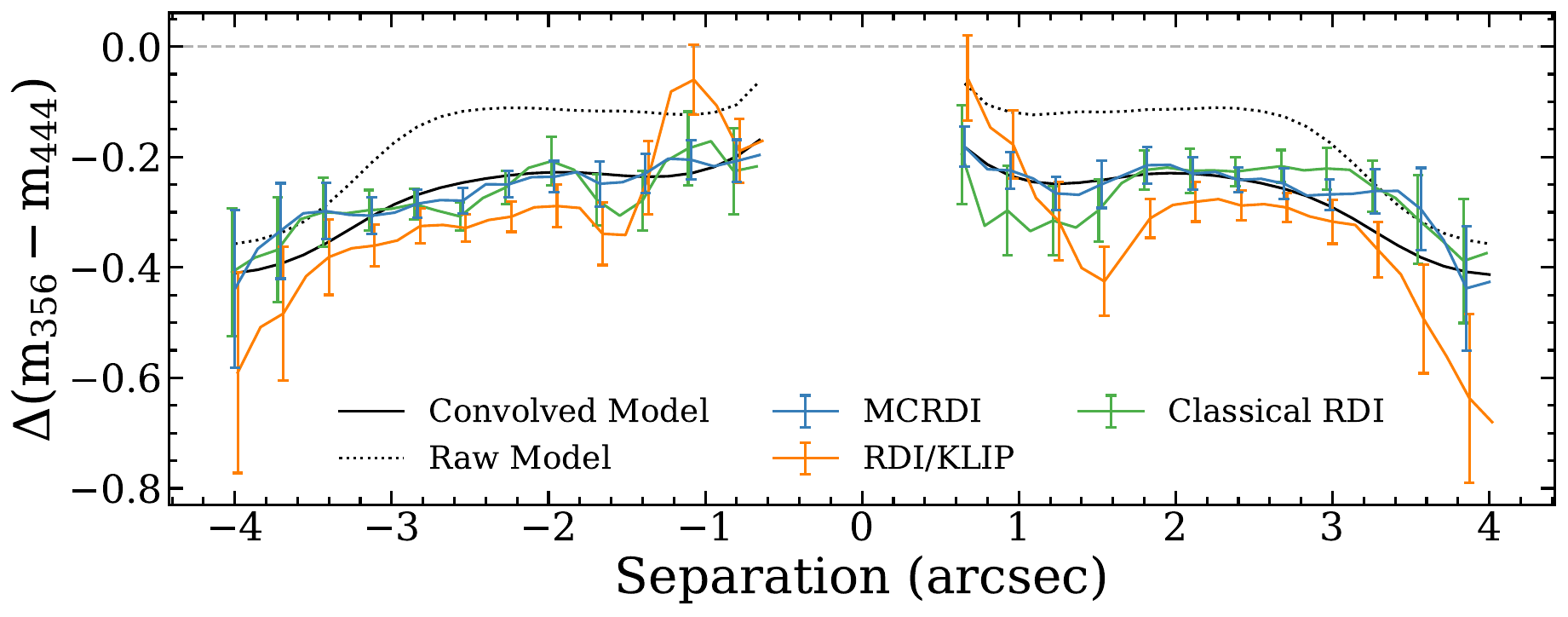}
\caption{Measurements of AU Mic's disk color as described in Section \ref{sec:sb}. Measurements for the MCRDI disk model (black) are included for both the convolved (solid) and unconvolved (dotted) model images to provide a qualitative estimate of the effect of the differing PSFs on the measured color. To improve readability, a) error bars are plotted for every other measurement and b) points for the three data reductions are offset slightly from one-another in the x-axis direction. \label{fig:disk_color}}
\end{figure*}

The uncertainties for disk color measurements assume zero uncertainty for the stellar fluxes, such that all color uncertainty is contributed by the uncertainty for the disk \ac{sb}. Disk color as a function of projected separation is presented in Figure \ref{fig:disk_color}.

We remark that these measurements have not been corrected for the effects of the PSF and coronagraph beyond the typical JWST flux calibration strategy. As such, a brightness measurement at a given position may be contaminated significantly by the wings of the PSFs for other parts of the disk \citep[e.g., as in the Early Release Science observations of HD 141569 A with JWST MIRI;][]{Hinkley2023}. 

\subsection{Discussion of Disk Surface Brightness and Color}
The surface brightness profiles presented in Figures \ref{fig:sb} and \ref{fig:log_log_sb} manifest with similar shapes to those presented in shorter wavelength scattered-light studies — albeit smoothed by the wider 3$-$5$\rm \,\micron$ \ac{psf}. Most visible in Figure \ref{fig:log_log_sb}: the profiles show a modest slope from $r\sim$$1-3\arcsec$, which steepens beyond $r\sim$$3-4\arcsec$ (the ring's peak density radius) — as is noted in many prior studies of the system \citep[e.g.,][]{Liu2004a, Krist2005, Fitzgerald2007}. While the brightness profiles show a marked flattening within $\sim$$1\arcsec$, we note that this region is significantly affected by the coronagraph — such that much or all of this apparent change is likely instrumental rather than astrophysical (see Appendix \ref{app:instrumental_asymmetry} for additional context).

Taken at face value, \edit1{the measurements presented in Figures \ref{fig:log_log_sb} and \ref{fig:sb_asymmetry}} seem to suggest that the NW side of the disk is brighter for $r \lesssim 2\arcsec$ — \edit1{qualitatively} consistent with measurements at shorter wavelengths \citep[e.g.,][]{Fitzgerald2007, Schneider2014, Boccaletti2018}. However, this apparent asymmetry is significantly smaller than the \edit1{asymmetry in HST/STIS imagery of the disk reported by \citet{Schneider2014}}. \edit1{For comparison, the asymmetry for the exponential fit to the \ac{sb} measurements for the NW and SE extents of the disk, provided in Figure 39 of \citet{Schneider2014}, is included in Figure \ref{fig:sb_asymmetry}.}
Additionally, comparison with the measurements of the truly symmetric MCRDI disk model, also displayed in Figure \ref{fig:sb_asymmetry}, reveals that the two sets of measurements are generally consistent with one another at small separations\footnote{A low significance deviation at $r\sim 1\farcs25$ is apparent in the F356W profile of Figure \ref{fig:sb_asymmetry}. However, given the lack of recurrence in F444W, we conclude that it is not likely to be astrophysical — or, if astrophysical, is much brighter in F356W and thus unlikely to be a companion (which should be much brighter in F444W).}. This suggests, instead, that much of the apparent asymmetry is induced by instrumental effects. Additional testing summarized in Appendix \ref{app:instrumental_asymmetry} shows that the misalignment of the coronagraph (see Section \ref{sec:imreg}) can explain the majority of the asymmetry at small separations, with some additional asymmetry resulting from the asymmetry of the NIRCam PSF itself. \edit1{The lower panel of Figure \ref{fig:sb_asymmetry} shows the result when correcting the on-sky asymmetry measurements for the asymmetry induced in the model. Measuring the overall surface brightness of the inner disk using a large rectangular aperture spanning $0\farcs63$ to $2\arcsec$ in the major axis direction and $-0\farcs5$ to $0\farcs5$ in the minor axis direction — and correcting for instrumentally-induced asymmetry in the same manner — yields overall inner disk asymmetry measurements of $-0.007\pm0.019$ mags for F356W and $-0.014\pm0.023$ mags for F444W.\footnote{The uncertainties for the corrected asymmetry measurements do not account for additional (but difficult to quantify) uncertainties introduced by this correction strategy — such as the accuracy of the determined coronagraph misalignment, the accuracy of the synthetic PSFs, or the accuracy of the disk model itself — and thus should be considered optimistic estimates.}} Overall, it appears likely that the enhanced brightness of the NW disk observed at shorter wavelengths is absent or significantly diminished at these wavelengths. 

 At larger separations, a seemingly localized increase in brightness on the SE side for both filters pushes the SE brightness above that of the NW at $r\sim2\farcs{}5$, with a second lower significance enhancement at $r\sim3\farcs{}25$. Notably, the former roughly coincides with the projected location of the \texttt{SE4} feature from \citet{Boccaletti2018} for the epoch of these data.  While these data do not generally have sufficient spatial resolution to study these features in detail, this may be suggestive that NIRCam can identify them through such brightness enhancements. The recurrence of this feature in both filters lends further credence to this notion.

The colors measured from the classical RDI and MCRDI reductions (Figure \ref{fig:disk_color}) are generally consistent — indicating an approximately flat or slowly-changing blue color of approximately $-0.3$ mags over the field of view, with a low significance blue dip near $4\arcsec$. The anticipated small-separation stellar residuals for the classical RDI data manifest here as noisy/erratic measurements within $\sim 2\arcsec$. A key risk for using oversubtracted results to study disks is highlighted by the RDI/KLIP disk color profile in Figure \ref{fig:disk_color}, which shows measurements deviating significantly from those of classical RDI and MCRDI alongside suggestive sharp radial trends. Notably, a) the disk flux lost to oversubtraction is the projection of the underlying disk image onto the reference images and b) the width of the stellar diffraction pattern scales linearly with the observing wavelength while the true distribution of disk flux very likely does not. As a result, oversubtraction typically varies both spatially and spectrally, and so will tend to induce varying offsets in measurements of disk color.

Comparison of the color profiles for the raw and convolved models indicates a color offset introduced by the convolution as a result of the differences between the two filters' PSFs. In galactic astronomy, when measuring colors and color gradients of galaxies, this effect is often circumvented by convolving each filter's images with the PSF of the other filter before measuring surface brightness profiles \citep[``cross-convolution"; e.g.,][]{denBrok2011}. This produces images at differing wavelengths in which flux from sources is smeared comparably — but at the cost of degrading the effective spatial resolution of the data. Applying this approach to our MCRDI-reduced images, the result is much more similar to the measurements for the raw (unconvolved) model, but shows a more uniform and gentle slope: with disk color values spanning roughly $-0.1$ to $-0.25$ mags from $1\arcsec$ to $4 \arcsec$. Deconvolution techniques, such as Richardson–Lucy deconvolution \citep{Richardson1972, Lucy1974}, may provide a solution that avoids degrading the spatial resolution of the data. However, in our case, these methods are significantly challenged by both the spatial variations of the NIRCam PSF and the use of a coronagraph. Ultimately, we leave a more detailed assessment of possible solutions for future work, and provide the current measurements as a preliminary assessment.

\edit1{Though many prior scattered-light studies of the disk also found a blue disk color \citep[e.g.,][]{Krist2005, Fitzgerald2007, Lomax2018}, we emphasize that this is the first time the disk's 3--5$\, \micron$ color has been measured. The NIRCam color probes a distinct region of the scattered light spectrum and so is sensitive to different properties of the disk material. For example, optical color is much more effective for assessing the minimum grain size for AU Mic's disk \citep[$\sim$$\, 0.2 \, \micron$;][]{Arnold2022}. Assuming a composition of standard astronomical silicates and a grain size distribution otherwise as the best-fit ADP solution in \citet{Arnold2022}: the difference between a minimum grain size of 0.1 and 0.2 $\micron$ is a $>$$1$ magnitude change in the 0.4 to 0.9 $\micron$ color, while the 3.5 to 4.4 $\micron$ color is negligibly affected (changing on the order of millimags). On the other hand, the 3--5$\, \micron$ regime is rich with scattered light spectral features \citep[e.g., tholins, $\rm C O_2$ ice, $\rm H_2 O$ ice;][]{Rodigas2015, Chen2019, Kim2019}, while the optical regime is predominantly featureless. As such, it is possible — if not likely — that the blue NIRCam color is the result of a distinct mechanism from the blue color at shorter wavelengths. As illustrated in \citet{Arnold2022}, assessing dust composition and size distribution in debris disks via multi-wavelength observations requires particular care.} The extensive degeneracies that exist between various compositional mixes and size distributions mean that identifying a single strong solution is insufficient. Rather, it is necessary to conduct a thorough investigation of all reasonable solutions, such that some subset of solutions that are consistent with the data can be identified. With consideration for this, we reserve further exploration of these details for a follow-up study \edit1{that considers these NIRCam measurements alongside those of prior studies}.

\section{Vertical Disk Structure}\label{sec:vertical_structure}
\edit1{The projected vertical size of the disk (i.e., its apparent width in the minor axis direction) is primarily a function of the disk's orientation, vertical dust distribution, and radial dust distribution (assuming $i \neq 90\degr$). As such, detailed modeling is necessary before this observable can be distilled to more physically valuable information regarding the disk's vertical structure, and thus to a better understanding of the mechanisms shaping the disk \citep[e.g.,][]{Daley2019}. Notably, the disk's projected vertical size must be resolved for such an investigation to be meaningful. Whether this is the case for the NIRCam data is not immediately clear, as t}he spatial resolution is comparable to the \edit1{projected} vertical scale expected for the disk based on prior observations. \edit1{For example, using HST/ACS imagery of the disk (FWHM$\,\sim\,$$0\farcs063$ at F606W), \citet{Krist2005} measure a projected disk FWHM as small as $\sim\,$$0\farcs22$ (occurring at a projected separation of $\sim\,$$1\farcs5$), while the FWHMs of the F356W and F444W filters are $0\farcs14$ and $0\farcs18$ respectively.} 

\edit1{To assess whether the projected vertical extent of the disk is resolved in NIRCam imagery, we proceed as follows. Along several vertical slices (perpendicular to the major axis), we measure the disk's surface brightness in the F356W MCRDI reduction.} These measurements are made using rectangular apertures with width 3 pixels ($0\farcs189$) and height 1.5 pixels ($0\farcs095$) — with each profile then being normalized to have the same peak value. \edit1{For comparison, we make like-measurements of a ``thin model" of the disk having negligible projected vertical size ($h_0 / r_0 = 10^{-5}$, $\alpha_{in} = 10$, $\alpha_{out} = -10$, $\gamma = 2$\edit1{; otherwise as given in Table \ref{tab:model_params}})\footnote{\edit1{See Appendix \ref{app:mcrdi_models} for descriptions of these parameters}} and convolved with a) coronagraphic images from \texttt{WebbPSF\_ext}, and b) an empirical field PSF\footnote{\texttt{WebbPSF} and \texttt{WebbPSF\_ext} do not account for a number of detector-level effects that can result in underestimation of the PSF width.}} extracted from a bright background source in the near-contemporaneous GTO 1184 observations of TYC 5899. \edit1{If the projected vertical size of AU Mic's disk is resolved in the NIRCam data, a PSF-convolved image for a disk much thinner than is supported by prior observations should yield substantially narrower vertical profiles than the on-sky data. Likewise, if the measurements of the aforementioned thin model yield profiles of comparable width to the on-sky data, it would mean that the projected vertical size is not resolved.
}

\edit1{Alongside these measurements, we also provide profiles for the instrumental \ac{psf} for reference. A}t each separation, we used \texttt{WebbPSF\_ext} to generate a \ac{psf} at the location of the disk major axis for each roll-angle and then derotated and combined the images as we did for the data. This results in a sequence-averaged \ac{psf} having FWHM of $\sim 0\farcs18$ in vertical profiles measured using the aforementioned aperture. 

\begin{figure}
\centering
\includegraphics[width=0.45\textwidth]{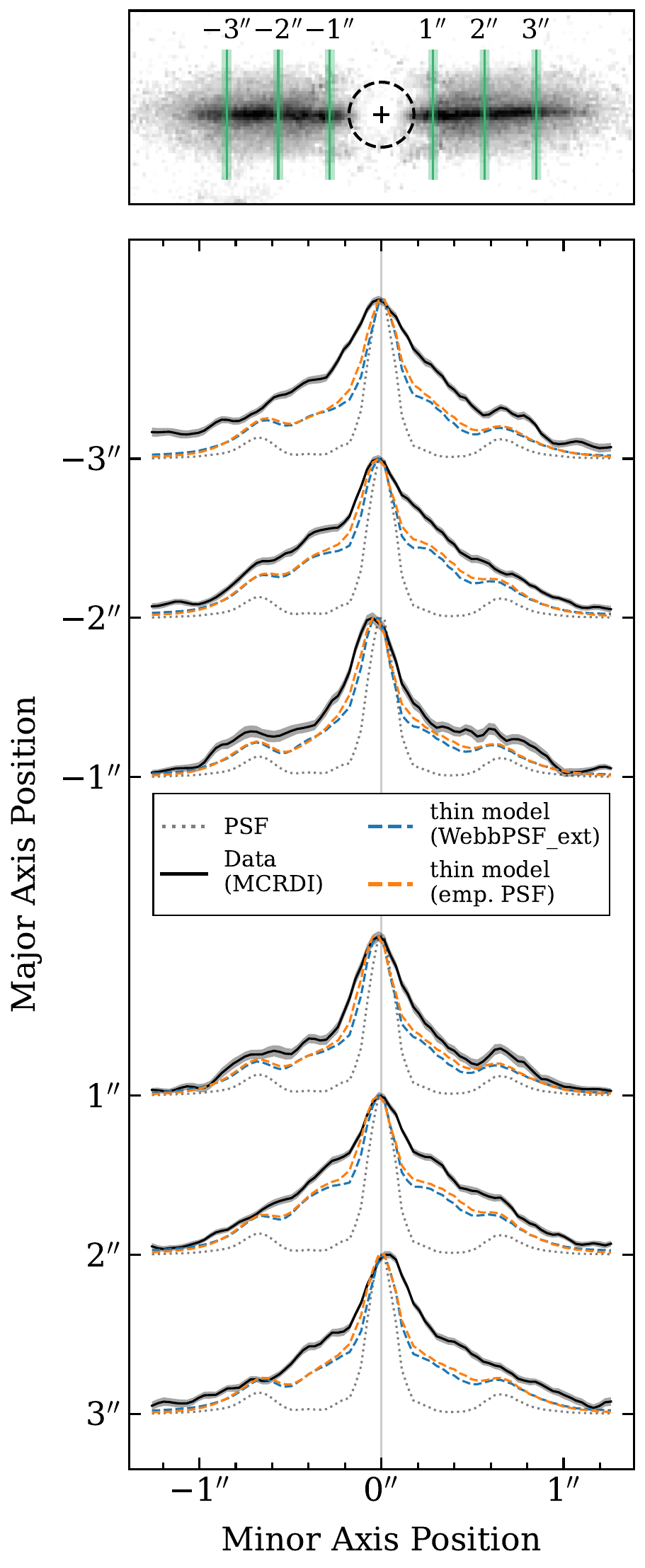}
\caption{Normalized vertical brightness profiles for the F356W images described in Section \ref{sec:vertical_structure}. The profiles are normalized to have the same peak value and are offset along the y-axis according to the major axis position of the slice. Shaded gray regions indicate the $1\sigma$ uncertainties on the measurements for the MCRDI profiles. The top panel shows the locations and widths of the six slices in the context of the MCRDI image\edit1{ (which has been multiplied by the projected radial separation for this visualization only)}. \label{fig:vertical_slices}}
\end{figure}

\edit1{The measured vertical brightness profiles are presented in Figure \ref{fig:vertical_slices}.} Uncertainties ($1\sigma$) for the measurements of the data are depicted as shaded gray regions and were computed following the same procedure outlined in Section \ref{sec:sb}. 

\subsection{Discussion of Vertical Structure}
 As evident from comparison of the profiles for the \ac{psf} and the thin disk model using \texttt{WebbPSF\_ext} for convolution: even a disk of unresolved vertical scale will manifest with much greater width than the \ac{psf} as a result of the extended nature of the disk flux and the shape of the NIRCam \ac{psf}. As such, comparison with profiles of the \ac{psf} alone is insufficient for assessing whether the vertical extent of the disk is resolved. However, the profiles for the data are significantly wider than the profiles for the thin model — for both the \texttt{WebbPSF\_ext} and empirical PSF versions. This suggests that we are seeing disk signal that is at least marginally resolved in the minor axis direction. \edit1{As further evidence of this, we note that our best-fit MCRDI disk model  appears consistent with the projected vertical size of the disk reported in \citet{Krist2005}. At $2\arcsec$, \citet{Krist2005} report a projected disk FWHM of $0\farcs25$ for HST/ACS F606W imagery. Convolving our raw F356W MCRDI disk model with a Gaussian kernel with FWHM$\,=0\farcs063$ to approximate the F606W PSF, we measure a very similar projected disk FWHM value of $0\farcs24$ at this separation.}

\section{Conclusions and Future Work}

We have presented high-contrast coronagraphic imaging of the AU Mic debris disk system from JWST/NIRCam. Our key findings are summarized hereafter.

\begin{enumerate}

\item The disk is unambiguously recovered in both F356W ($3.563 \, \micron$) and F444W ($4.421 \, \micron$) from separations as small as $\sim 0\farcs3$ (well inside of the \ac{iwa}) to separations as large as $\sim5\arcsec$. These detections mark the first images of the disk at 3--5$\, \micron$.

\item Using the model-constrained RDI (MCRDI) technique for the removal of starlight, final images are free of both the systematic oversubtraction that plagues RDI/KLIP or RDI/LOCI products and the significant small-separation stellar residuals apparent in classical RDI products.

\item No companions were identified, but analysis indicates that the data were capable of uncovering planets as small as $\sim 0.1 \; {\rm M_{J}}$ with $5\sigma$ confidence beyond $\sim 2\arcsec$ ($\sim 20$ au). These deep constraints on the presence of wide-orbit, massive planets are unique for AU Mic and provide relevant context for understanding planet formation and evolution in a system with a dynamic debris disk and a compact, multi-planet system at $<<$1 au separations. 

\item \edit1{The significant brightness asymmetry favoring the northwestern side of the inner disk at shorter wavelengths is not evident at 3--5$\, \micron$ — with a 1$\sigma$ upper limit of $\sim\,$$0.04$ mags on any asymmetry (versus the best-fit model with asymmetry of $\sim\,$$0.2$ mags from \citealt{Schneider2014}).}

\item In both filters, we see evidence of one or more localized brightness enhancements to the southeast that may correspond to previously-identified fast-moving disk features \citep[e.g.,][]{Boccaletti2015}.

\item A blue disk color of roughly $-0.3$ mags is measured between F356W and F444W — likely corresponding to a color closer to $-0.2$ mags when PSF effects are considered.

\item The \edit1{projected} vertical size of the disk is slightly resolved in these data, showing an apparent FWHM of $0\farcs42$ at $1\arcsec$ in F356W, compared to $0\farcs25 - 0\farcs28$ for a disk model having infinitesimal vertical size before convolution with the instrumental PSF.
\end{enumerate}

The high-significance detection of AU Mic's disk at these wavelengths demonstrates the unique suitability of JWST and NIRCam for studying this system — and motivates further studies facilitated by JWST's wavelength coverage. Follow-up observations in additional \edit1{NIRCam} filters would permit the study of the presence and location of ices in the system \edit1{by further probing the 2–5 $\mu m$ scattered-light spectral features of ices alongside the surrounding continuum \citep[e.g.,][]{Honda2009, Debes2013, Tazaki2021, Betti2022} — thus testing the inferred presence of water ice based on polarimetric imagery from HST/ACS in \citet{Graham2007}}. \edit1{As the theoretical ice line for AU Mic falls at $\sim$$\,2$ au \citep{Schuppler2015}, ices should be observed throughout the bulk of the resolvable disk.} With the planned addition of dual-channel coronagraphic imaging for NIRCam, \edit1{the presence of ices could be tested} with just one additional sequence using, e.g., F210M and F300M. 

\edit1{As demonstrated in \citet{Rodigas2015}: the addition of thermal IR data can substantially alter the interpretation of a disk compared to scattered-light data alone.} Following from \citet{Gaspar2023}, who reported mid-infrared imaging of the Fomalhaut debris disk with JWST's Mid-Infrared Instrument (MIRI), imaging of AU Mic with MIRI would likewise probe AU Mic's disk in the thermal regime — further constraining its composition and helping to clarify the factors shaping it.

Re-observation of the system with NIRCam in the F444W filter would serve to place stronger constraints on the outer planets that might still remain hidden. In many cases, projected locations very near the coronagraph center are the only orbital positions that cannot be ruled out with high confidence from the existing data. A second epoch even just a year after the first would provide a sufficient baseline for the few degrees of orbital motion needed for many of these would-be planets to become detectable.

Overall, the results presented herein highlight the strength of JWST/NIRCam for studying circumstellar disk systems. While ground-based facilities observing at shorter wavelengths may achieve better spatial resolutions and reach higher contrasts, JWST is the only observatory capable of studying many compelling targets in the $\sim$ 2–5$\,\micron$ regime. As these wavelengths coincide with extremely favorable planet contrasts and span numerous notable scattered-light spectral features, this capability has considerable scientific value.

\acknowledgements
We thank our referee, whose comments helped us to improve both the content and clarity of this manuscript.

We acknowledge the decades of immense effort that enabled the successful launch and commissioning of the JWST; these results were possible only through the concerted determination of thousands of people involved in the JWST mission. In particular, we offer gratitude to a number of individuals who enabled this study through contributions to either the 2002 NIRCam instrument proposal or to the development and commissioning of the NIRCam instrument: Martha Boyer, Daniel Eisenstein, Klaus Hodapp, Scott Horner, Doug Kelly, Don McCarthy, Karl Misselt, and George Rieke. We are grateful for support from NASA through the JWST NIRCam project, contract number NAS5-02105 (M. Rieke, University of Arizona, PI). 

The JWST data presented in this paper were obtained from the Mikulski Archive for Space Telescopes (MAST) at the Space Telescope Science Institute. The specific observations analyzed can be accessed via \dataset[https://doi.org/10.17909/x9hs-pr32]{https://doi.org/10.17909/x9hs-pr32}. STScI is operated by the Association of Universities for Research in Astronomy, Inc., under NASA contract NAS5–26555. Support to MAST for these data is provided by the NASA Office of Space Science via grant NAG5–7584 and by other grants and contracts.

This publication makes use of data products from the Wide-field Infrared Survey Explorer, which is a joint project of the University of California, Los Angeles, and the Jet Propulsion Laboratory/California Institute of Technology, funded by the National Aeronautics and Space Administration.

K. Lawson’s research was supported by an appointment to the NASA Postdoctoral Program at the NASA–Goddard Space Flight Center, administered by Oak Ridge Associated Universities under contract with NASA.

E. Bogat's work was supported by a grant from the Seller's Exoplanet Environments Collaboration (SEEC) at NASA GSFC, administered through NASA's Internal Scientist Funding Model (ISFM). 

\software{Matplotlib \citep{matplotlib2007, matplotlib2021},
NumPy \citep{numpy2020},
SciPy \citep{scipy2020},
Astropy \citep{astropy2013, astropy2018},
CuPy \citep{cupy2017},
LMFIT \citep{Newville2014},
WebbPSF \citep{Perrin2014},
WebbPSF\_ext \citep{Leisenring2021},
SpaceKLIP \citep{Kammerer2022},
Vortex Image Processing \citep{Gonzalez2017}}

\appendix
\section{MCRDI Disk Models and Optimization}\label{app:mcrdi_models}

To generate models for use in the model-constrained \ac{rdi} procedure, we utilize the \texttt{scattered\_light\_disk} module of the \texttt{Vortex Image Processing} \citep[\texttt{VIP};][]{Gonzalez2017} Python package, which introduces a ``lite" version of the GRaTer disk modeling code \citep{Augereau1999} which assumes an optically thin disk dominated by single scattering. Since the goal of our application is merely to superficially simulate the distribution of light from the disk for the purpose of preventing \ac{rdi} oversubtraction, the physical validity of this assumption is inconsequential. Though the AU Mic disk is famous for structure and asymmetries observed at shorter wavelengths, inspection of the other NIRCam data reductions suggested that any such features are much less pronounced or absent in these data (due to wavelength dependence or spatial resolution). Moreover: to eliminate oversubtraction with MCRDI, it is not necessary to perfectly reproduce every nuance of the distribution of \ac{css} in the data. Rather, the \ac{css} can be over-estimated in some areas and under-estimated in others — so long as the two balance in the least-squares construction of the model of the stellar diffraction pattern (i.e., such that the residuals between the true \ac{css} and the estimated \ac{css} yield a negligible projection onto the reference images). As such, we adopt a simple axisymmetric ring-like disk geometry.

Using \texttt{VIP}, we model the disk as a function of nine parameters and with a \ac{spf} that is the linear combination of two Henyey-Greenstein (H-G) SPFs \citep{Henyey1941}. The models assume linear flaring ($\beta = 1$) and a Gaussian vertical density distribution ($\gamma = 2$). The varied parameters include:

\begin{itemize}
    \item inclination (incl; degrees)
    \item position angle ($\rm PA$; degrees)
    \item fiducial radius ($\rm r_0$; au)
    \item the ratio of scale height to radius ($\rm h_0 / r_0$)
    \item the density power-law indices interior and exterior to the fiducial radius ($\alpha_{in}$ and $\alpha_{out}$, respectively)
    \item the H-G asymmetry parameters, $g_1$ and $g_2$, and the weight for the first SPF ($w_1$; with $w_2 = 1 - w_1$)
\end{itemize}

We create each raw model oversampled by a factor of two relative to the data. We then rotate the model to the appropriate roll angles for the data and convolve it with synthetic NIRCam coronagraphic images — sampled from an array of detector positions. These images are generated with the \texttt{WebbPSF\_ext} package and using the \ac{opd} map measured closest in time to the observations.

In lieu of precise prior knowledge of the disk's spectrum (which will affect the resulting diffraction pattern), we instead adopt a synthetic spectrum matching the spectral type of the parent star. Effectively, this assumes that scattered light is dominant at these wavelengths and that this scattering lacks wavelength dependence (i.e., ``gray scattering"). Though not precisely accurate, this approximation is sufficient for the purpose of the desired superficial disk estimate.

At this point, the synthetic coronagraphic images to be used for convolution could be normalized as in Section \ref{sec:companion_limits}: by normalizing to the transmission of the coronagraph at each sampled position. This would permit a convolved image that effectively samples the changes to the diffraction pattern's morphology and to the coronagraph transmission at the resolution of the sampled grid. To permit finer sampling of the transmission, we instead normalize each of the synthetic coronagraph images to sum to one, but multiply each model image with a coronagraph transmission map just before convolution — resulting in effective transmission sampling at the resolution of the oversampled pixels. We remark that this significantly improves the consistency of the final disk models with the data in this case; we strongly recommended this strategy for any convolution scenarios where circumstellar signal is present near the \ac{iwa}.

\begin{figure*}
\centering
\includegraphics[width=0.98\textwidth]{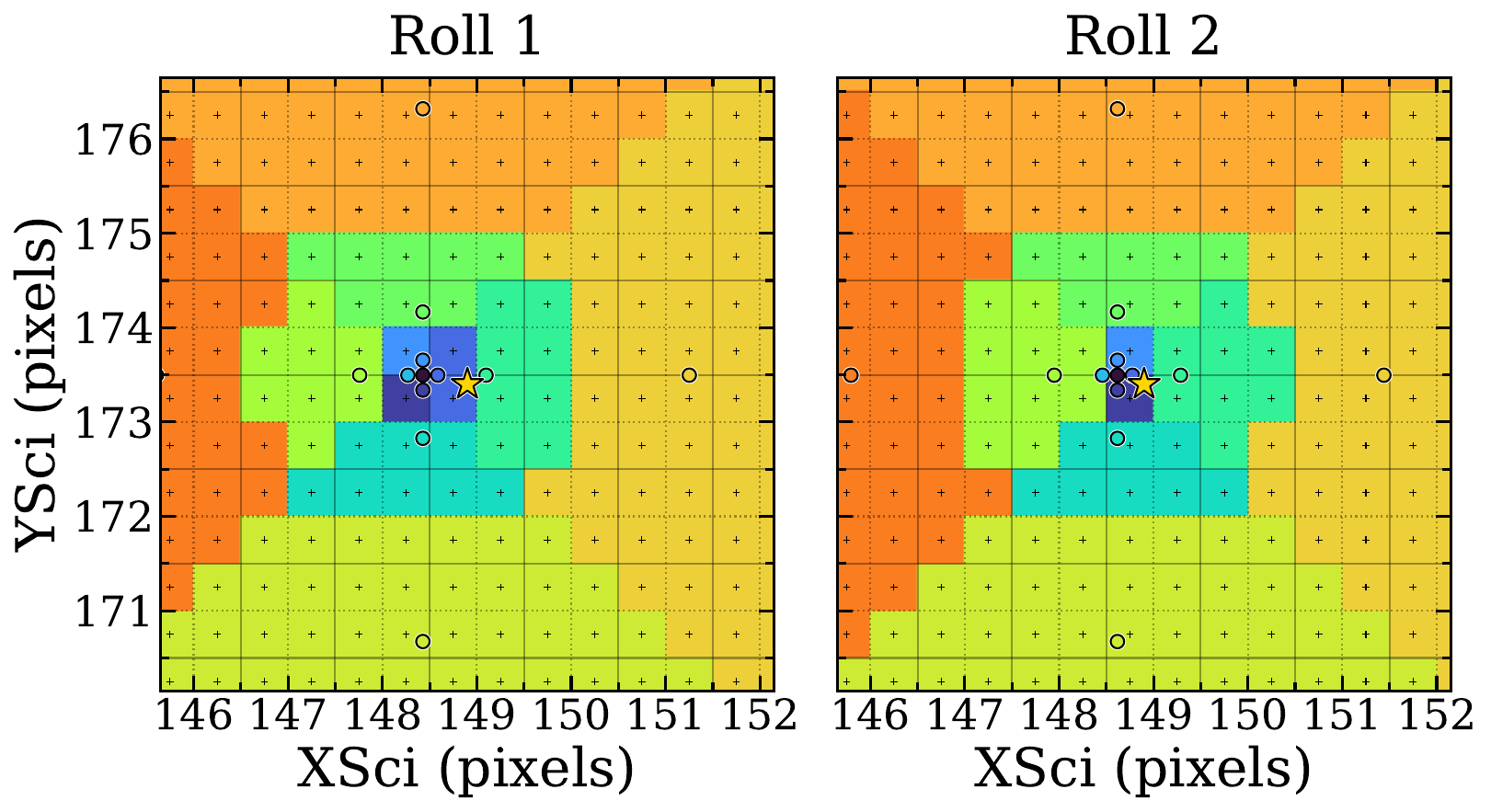}
\caption{For both rolls of the F356W AU Mic observations: a map showing the vicinity of the star (gold star marker) with each oversampled pixel colored according to the coronagraphic image sample (colored round markers) with which it is paired for model convolution. Solid vertical and horizontal lines denote the boundaries of detector-sampled pixels, while dotted lines delineate the oversampled pixel boundaries. Note: not all of the \texttt{WebbPSF\_ext} samples in the grid are ultimately matched to any pixels. \label{fig:nircam_psf_grid}}
\end{figure*}

We then create the convolved model image for each roll angle by convolving every pixel of the rotated input model with the nearest sample from the grid in polar coordinates (see Figure \ref{fig:nircam_psf_grid}). For each roll angle, we consider the previously computed offset of the coronagraph center from the star for both the coronagraphic transmission and for matching pixels with samples from \texttt{WebbPSF\_ext}. Once convolution is completed, we resample the resulting images to match the sampling of the data. 

Once a convolved model sequence is created for a particular set of input disk parameters, we forward model it for our standard LOCI \ac{rdi} reduction in the typical manner \citep[e.g.,][]{Currie2019} and compare it with the observed LOCI RDI result within a region of interest. This approach is mathematically identical to the direct optimization of a \ac{css} estimate using MCRDI\footnote{That is, computing the RDI residuals after first subtracting the \ac{css} estimate from the data; see \citep{Lawson2022} for more information regarding this equivalency.}, but allows for post facto analytic rescaling of the model's brightness to minimize residuals with the data — ultimately eliminating the need to vary the brightness of the disk alongside the other parameters\footnote{Since oversubtraction is entirely the result of the presence of the disk itself, the brightness of the forward modeled result scales linearly with that of the input. For example, if $I_M^\prime$ is the forward-modeled result for an input disk model $I_M$, then $s  I_M^\prime$ will be the result for the input model $s  I_M$ for any scalar constant $s$. The value of $s$ producing the best fit can therefore be computed after forward modeling.}. The region of interest includes pixels with either $r < 35$ pixels or falling within a stellocentric rectangular region rotated to the approximate position angle of the disk \citep[PA $= 128 \fdg 48$;][]{Vizgan2022} with a width of 12 pixels and a length of 94 pixels (where the detection in the LOCI RDI images begins to wane) — but excluding the region within 3 pixels of the star (where small differences in coronagraph alignment can produce significant stellar residuals that might otherwise impact disk model fitting). This region of interest is intended to include the majority of the disk flux while also including the inner background region where oversubtraction is most evident in the unconstrained LOCI result. We note that the region falling beyond the optimization region used for RDI subtraction has no effect on the amount of oversubtraction and thus does not need to be considered for the purpose of eliminating oversubtraction. However, as we needed a model that was superficially accurate at larger separations for use in Sections \ref{sec:companion_limits} and \ref{sec:sb}, we chose to extend the region of interest as described to cover both needs.

To explore the resulting disk model parameter-space, we use the differential evolution algorithm \citep{Storn1997} — a genetic algorithm for global optimization which performs well in degenerate and/or multi-modal parameter spaces and which typically identifies a strong solution in only a few thousand samples \citep[e.g.,][]{Lawson2020}. For optimization, all of the aforementioned parameters are varied in wide ranges — with the exception of position angle, which is fixed to $128 \fdg 48$. The model for each of the two filters was optimized separately with this approach — except that the F444W model was restricted to the best-fit inclination of the F356W model.

After the optimal model is identified in this manner, the scaled and convolved input model is used to perform MCRDI on the data to achieve the final MCRDI result. The MCRDI result, the best-fit model estimate, and the residuals for each filter are shown in Figure \ref{fig:mcrdi_models}. 

In Table \ref{tab:model_params}, we provide the best-fitting model parameters. However, we emphasize that these are included purely for the sake of repeatability. These values were not optimized to be physically meaningful, but rather to superficially emulate the distribution of disk flux following convolution. These values should not be used other than to reproduce the MCRDI reductions for these or very similar data. 

\begin{deluxetable*}{@{\extracolsep{5pt}}cccccccccc}
    \tablewidth{0pt}
    \tablecaption{MCRDI Disk Model Parameters}
    \tablehead{
    \colhead{Filter} & \colhead{Incl (deg)} & \colhead{$\rm PA$ (deg)\tablenotemark{a}} & \colhead{$\rm r_0$ (au)} & \colhead{$\rm h_0 / r_0$} & \colhead{$\alpha_{in}$} & \colhead{$\alpha_{out}$} & \colhead{$g_1$} & \colhead{$g_2$} & \colhead{$w_1$}
    }
    \startdata
    F356W & 89.20 & 128.48 & 33.49 & 0.038 & 2.82 & -4.92 & 0.95 & 0.46 & 0.44 \\
    F444W & 89.20 & 128.48 & 30.26 & 0.039 & 3.39 & -4.85 & 0.98 & 0.47 & 0.65 \\
    \enddata
    \tablecomments{Disk model parameter values used in the Model-Constrained RDI reductions presented herein. These are provided solely for the sake of reproducibility and should not be considered robust measurements of disk parameters. See Appendix \ref{app:mcrdi_models} for details. \tablenotetext{a}{The value passed into \texttt{VIP} is $\rm PA - 180\degr$ due to a difference of convention.}}\label{tab:model_params}
\end{deluxetable*}

\section{Instrumental Asymmetry Effects}\label{app:instrumental_asymmetry}

\begin{figure*}
\centering
\includegraphics[width=0.98\textwidth]{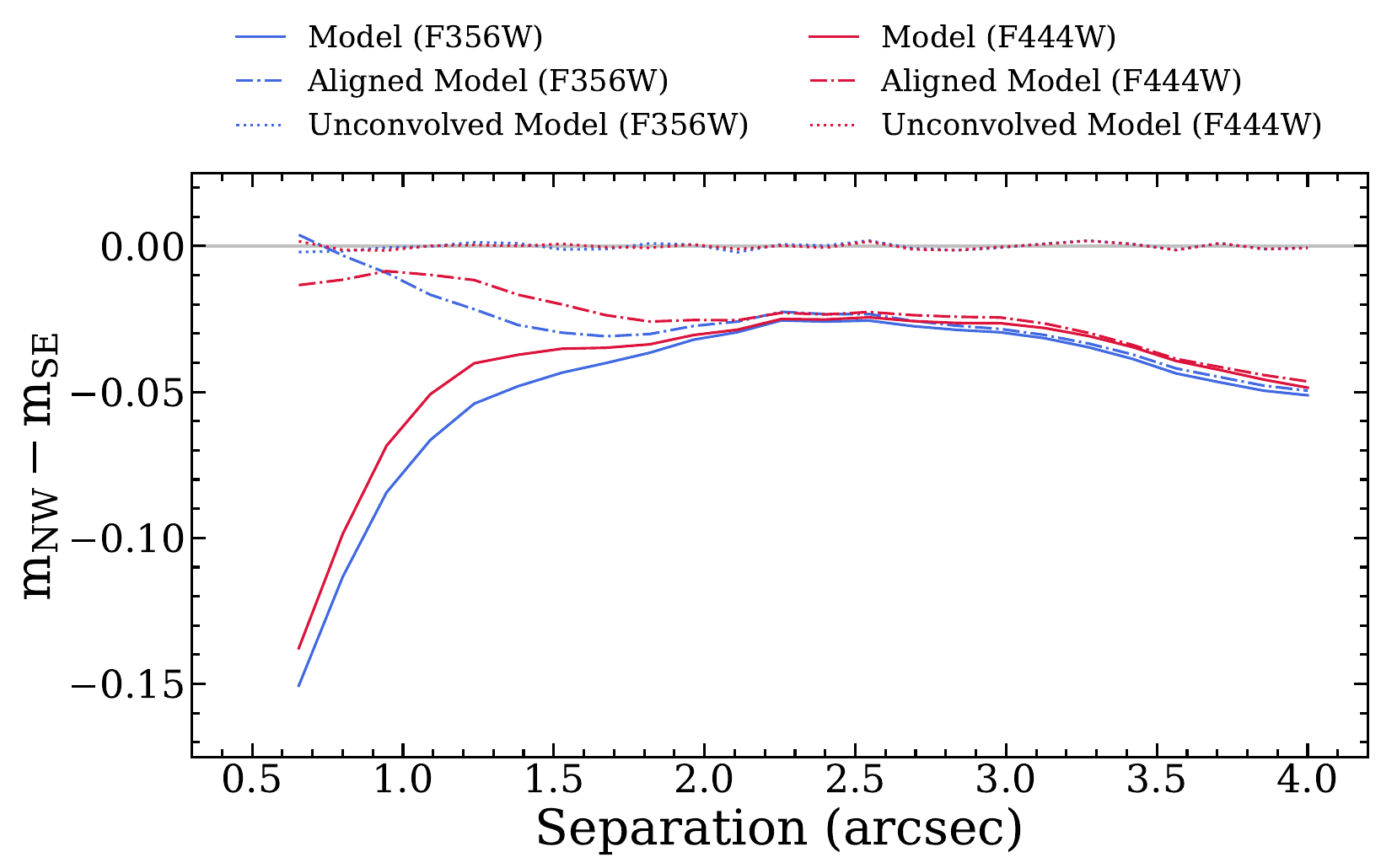}
\caption{As Figure \ref{fig:sb_asymmetry}, but for measurements of differing disk models and with color now encoding the filter (F356W in blue and F444W in red). Solid lines correspond to measurements of the convolved MCRDI model (as the black lines in Figure \ref{fig:sb_asymmetry}), the dash-dotted lines are for the same model but assuming the star and coronagraph are perfectly aligned for convolution, and the dotted lines are for the model before convolution.
\label{fig:model_asymmetry}}
\end{figure*}

To diagnose the cause of the induced asymmetry for the otherwise symmetric input MCRDI model discussed in Section \ref{sec:sb}, we make additional asymmetry measurements using the same input model but performing convolution assuming perfect alignment between the star and coronagraph for both rolls. Figure \ref{fig:model_asymmetry} compares these results with those of the nominal convolved model and with the unconvolved input model. This reveals that the majority of the small-separation asymmetry results from the misaligned coronagraph, which effectively blocks more of the disk on the SE side. The reach of this effect to separations as large as $\sim2 \arcsec$ is driven by the width of the NIRCam PSF combined with the sharp decline in disk brightness as separation increases --- resulting in a non-negligible contribution from the bright portion of the disk that is occulted to the southeast but not occulted to the northwest.

The remaining asymmetry for the ``aligned" model appears to result from the asymmetry of the PSF itself. Data simulated with parallactic angles offset from those of the data (but with the same roll-offset) show broadly varying induced asymmetries. These asymmetries were minimized when the major axis of the disk was aligned with the y-axis direction of the detector — where the PSF-core is more symmetrical — and never became significantly larger than the asymmetries induced at the observed parallactic angles.

\bibliography{refs}{}
\bibliographystyle{aasjournal}
\end{document}